\documentclass[a4paper,11pt]{article}
\pdfoutput=1 

\usepackage{jheppub} 

\usepackage[T1]{fontenc} 

\usepackage{graphicx}
\usepackage{epsfig}
\usepackage{rotating}
\usepackage{amssymb}
\usepackage{subfigure}
\usepackage{dsfont}
\usepackage{psfrag}
\usepackage{amsmath,euscript,array,mathrsfs}
\usepackage{bbold}
\usepackage{epsf}
\usepackage{tablefootnote}

\newcommand{\beq}{\begin{equation}}
\newcommand{\eeq}{\end{equation}}
\newcommand{\beqs}{\begin{eqnarray}}
\newcommand{\eeqs}{\end{eqnarray}}
\newcommand{\lsim}{\mathrel{\raisebox{-
.6ex}{$\stackrel{\textstyle<}{\sim}$}}}

\newcommand{\Tr}{{\rm Tr}}

\def\hbar{\hspace{0pt}\raisebox{1pt}{$-$} \hspace{-7pt} h}

\def\r{\rho}

\newcommand{\be}{\begin{equation}}
\newcommand{\ee}{\end{equation}}

\newcommand{\bea}{\begin{eqnarray}}
\newcommand{\eea}{\end{eqnarray}}
\newcommand{\nn}{\nonumber}

\def\lbldef#1#2{\expandafter\gdef\csname #1\endcsname {#2}}

\def\href#1#2{#2}

\newcommand{\ber}{\begin{eqnarray}}
\newcommand{\eer}{\end{eqnarray}}

\newcommand{\beqar}{\begin{eqnarray}}

\newcommand{\eeqar}{\end{eqnarray}}

\newcommand{\nonu}{\nonumber}

\newcommand{\dsl}
  {\kern.06em\hbox{\raise.15ex\hbox{$/$}\kern-.56em\hbox{$\partial$}}}

\newcommand{\eeqarr}{\end{eqnarray}}
\newcommand{\ZZ}{{\rm \kern 0.275em Z \kern -0.92em Z}\;}

\def\CC{{\mathchoice
{\rm C\mkern-8mu\vrule height1.45ex depth-.05ex
width.05em\mkern9mu\kern-.05em}
{\rm C\mkern-8mu\vrule height1.45ex depth-.05ex
width.05em\mkern9mu\kern-.05em}
{\rm C\mkern-8mu\vrule height1ex depth-.07ex
width.035em\mkern9mu\kern-.035em}
{\rm C\mkern-8mu\vrule height.65ex depth-.1ex
width.025em\mkern8mu\kern-.025em}}}

\def\RR{{\rm I\kern-1.6pt {\rm R}}}

\def\ZZ{{\rm Z}\kern-3.8pt {\rm Z} \kern2pt}
\def\IB{\relax{\rm I\kern-.18em B}}
\def\ID{\relax{\rm I\kern-.18em D}}
\def\II{\relax{\rm I\kern-.18em I}}
\def\IP{\relax{\rm I\kern-.18em P}}

\newcommand{\bear}{\begin{eqnarray}}
\newcommand{\eear}{\end{eqnarray}}

\def\r{\rho}                                     

\def\tt{\tilde{\theta}}

\def\6{\partial}

\def\bea{\begin{eqnarray}}
\def\eea{\end{eqnarray}}

\def\beqx{\begin{displaymath}}
\def\eeqx{\end{displaymath}}

\newcommand{\bmat}{\left(\begin{array}}
\newcommand{\emat}{\end{array}\right)}

\def\r{\rho}

\def\bo{{\raise-.3ex\hbox{\large$\Box$}}}               
\def\face{{\raise.2ex\hbox{$\displaystyle \bigodot$}\mskip-2.2mu \llap {$\ddot
        \smile$}}}                                   
\def\>{\rangle}                                      
\def\<{\langle}                                      

\def\leftrightarrowfill{$\mathsurround=0pt \mathord\leftarrow \mkern-6mu
        \cleaders\hbox{$\mkern-2mu \mathord- \mkern-2mu$}\hfill
        \mkern-6mu \mathord\rightarrow$}        
\def\dvec#1{\vbox{\ialign{##\crcr
        \leftrightarrowfill\crcr\noalign{\kern-1pt\nointerlineskip}
        $\hfil\displaystyle{#1}\hfil$\crcr}}}           
\def\Tr{{\rm Tr \,}}                                    

\def\-{\hphantom{-}}

\def\figdir{Figures/}

\newcommand\Eq[1]{Eq.~(\ref{eq:#1})}
\newcommand\Fig[1]{Fig.~\ref{fig:#1}}
\newcommand\Sec[1]{Sec.~\ref{sec:#1}}
\newcommand\Table[1]{Table~\ref{tab:#1}}

\newcommand\Tab[1]{Table~\ref{tab:#1}}

\title{Symmetry restoration at high-temperature in two-color and two-flavor lattice gauge theories}

\author[a,b,c]{Jong-Wan Lee,}

\affiliation[a]{Department of Physics, College of Science, Swansea University,
Singleton Park, SA2 8PP, Swansea, Wales, UK}

\affiliation[b]{Department of Physics, Pusan National University,
Busan 46241, Korea}

\affiliation[c]{Extreme Physics Institute, Pusan National University,
Busan 46241, Korea}

\author[a]{Biagio Lucini,}

\author[a]{ Maurizio Piai.}

\date{\today}

\abstract{
We consider the $SU(2)$ gauge theory with $N_f=2$ flavors of Dirac fundamental fermions.
We study the high-temperature behavior of the spectra of mesons, discretizing the theory
on anisotropic lattices, and measuring the two-point  correlation functions in the temporal direction
as well as screening masses in various channels. 
We identify the (pseudo-)critical temperature as the temperature at which the susceptibility associated 
with the Polyakov loop has a maximum. 
At high temperature both
the spin-1 and spin-0 sectors of the light meson spectra exhibit enhanced symmetry properties, indicating the restoration of both the 
global $SU(4)$ and the axial $U(1)_A$ symmetries of the model.
}

\begin{document}
\maketitle
\flushbottom


\newpage

\section{Introduction}
\label{Sec:intro}

We consider the $SU(2)$ gauge theory with $N_f=2$ flavors of Dirac fundamental fermions,
and study the finite-temperature behavior by using numerical methods based on  formulating the theory on  anisotropic lattices. 
The main purpose of this work is to collect evidence that the global symmetries of the model are implemented 
{\em \`a la Wigner} at 
high-temperature, where the condensate breaking global symmetry  is expected to melt
and the global symmetries to be linearly realized.

This model has been considered before in three different contexts, as it represents the prototype of
non-trivial gauge theory in which lattice numerical methods have concrete potential to provide 
useful information about the dynamics of the underlying theory.
First of all, it is a useful toy model for the study of generalizations of Quantum Chromo-Dynamics (QCD) at finite temperature $T$ and finite chemical potential $\mu$. 
One trivial reason for this is that the number of fundamental degrees of freedom is smaller than for two-flavor QCD,
making the numerical treatment easier.
Most importantly though, the fundamental representation of $SU(2)$ is pseudo-real, and hence there is no sign problem.
It is then possible to study the phase diagram of the model in the $(T,\mu)$-plane, and to apply numerical techniques
to extract its detailed structure.
For an incomplete list of useful references on the subject see~\cite{LatticeSU2}.  

A second context in which this model is  important is that of traditional technicolor (TC)~\cite{TC,TCreviews}. 
The choice of $SU(2)$ with $2$ fundamental Dirac fermions yields the minimal model such that 
one can embed the electro-weak $SU(2)_L\times U(1)_Y$ group of the Standard Model of particle physics (SM)
within the global symmetries of the matter field content. One expects spontaneous symmetry breaking to
arise dynamically at the scale $\Lambda$, hence providing a natural way to implement the Higgs mechanism 
for giving mass to the electroweak bosons within a fundamental theory.
Aside from the fact that, once more, the small number of degrees of freedom makes practical applications
amenable to numerical treatment, the fact that the field content is minimal also minimizes the potentially problematic
contributions to precision parameters such as the oblique $S$ and $T$ as defined by Peskin and Takeuchi~\cite{PT}, 
that on the basis of perturbative arguments one expects to grow
with $N_f$ and $N_c$, and that are not dynamically suppressed when one identifies $\Lambda$ with the electroweak scale $v_W\sim 246$ GeV. 
The dynamics preserves a custodial $SU(2)$ that further suppresses 
the $T$ parameter, as the underlying masses of the fermions vanish. 

The model has received some attention in a third context~\cite{sannino,SU4Sp4}, as a concrete realization of 
the idea of Higgs compositeness~\cite{compositeness}.
This is a quite distinct framework in respect to traditional TC.
The underlying dynamics is the same, being based upon a gauge theory with a given global symmetry,
for which one expects the formation of a non-trivial symmetry-breaking condensate.
Yet, one chooses to embed the electroweak gauge group into the global symmetry group of the theory 
in such a way that the fermion condensate does not break it.~\footnote{We ignore the problem of vacuum alignment~\cite{Peskin}.}
The long-distance behavior of the theory is hence captured by an 
Effective Field Theory (EFT) that includes the SM gauge theory,
supplemented by a set of light, composite  pseudo-Goldstone bosons arising at the scale $\Lambda$, a subset of which is interpreted as 
the Higgs doublet field. 

The gauging of the SM group explicitly breaks the global symmetries,
and hence provides a potential for the Higgs fields.
 Additional ingredients, not arising from the $SU(2)$ fundamental gauge theory,
are invoked in order to drive spontaneous 
symmetry breaking in the Higgs sector, which ultimately yields electro-weak symmetry breaking (EWSB)
at the scale $v_W \ll \Lambda$.
For example, one has to introduce a mechanism to give mass to the SM fermions, which requires coupling the Higgs field to the 
quarks and leptons.  It is well known that, as a byproduct of doing so, the theory yields radiative corrections to the Higgs potential due 
to loops of the top quark, naive estimates of which show that they can destabilize the minimum of the Higgs potential. 
In the following we will not discuss any of these points, related to realistic model-building in the electro-weak sector.

The reason why composite scenarios are viable within this model originates from the pseudo-real nature of the fundamental representation
of $SU(2)$. In particular, in the presence of two Dirac fermions, the global symmetry of the Lagrangian is enhanced from the 
$U(1)_A\times U(1)_B^t\times SU(2)^t_L\times SU(2)^t_R$ global symmetry of QCD and TC to a $U(1)_A\times SU(4)$ global symmetry,
and the condensate breaks it to the $Sp(4)$ subgroup.
Excluding for the time being the anomalous $U(1)_A$ from the discussion, this yields $5$ (pseudo-)Goldstone bosons,
that form a multiplet of the unbroken $Sp(4)\sim SO(5)$.
The gauging of $SU(2)_L\times U(1)_Y\subset SO(4) \subset Sp(4)$ splits the $5$ into a $4$ of $SO(4)$, 
which is identified with the Higgs doublet,
and an additional singlet, that may have important phenomenological implications. 

In this paper, we compute the masses of the composite (meson) states created and annihilated by 
operators of the form $\bar{Q} \Gamma Q$, with $\Gamma=1,\gamma_5\cdots$,
and discuss their dependence on temperature $T$.~\footnote{
We refer the reader to the works in~\cite{LatticeSU3Quenched,Cheng,LatticeSU3}: while these papers study $SU(3)$ gauge theory, 
some of their results and ideas play a role in the present paper.
} 
In particular we track how the  mass-splittings between parity partners
change by going to high-temperature.
In order to do so, we formulate the theory on  anisotropic lattices,
and use Monte Carlo methods to extract the spectral masses as a function of $T$.
We are looking for clear signals of the restoration in the thermal bath of the much larger global symmetry of the underlying theory.
This is the first step of a more ambitious and long-term program, which we envision will include also the 
study of the effects due to the presence of explicit symmetry-breaking terms,
in particular due to the chemical potential $\mu$, and to the weakly-coupled gauging of the SM electroweak group. 

The paper is organized as follows.
In Section~\ref{Sec:model} we describe the model and summarize effective field theory and symmetry arguments that 
play a role in the rest of the paper.
In Section~\ref{Sec:anisotropic_lattice} we describe the lattice set-up used in the numerical calculations,
particularly by explaining in details how the bare parameters are tuned in the presence of anisotropic lattices. 
In Section~\ref{Sec:finite_temp} we report our results, which we critically discuss in Section~\ref{Sec:discussion}.
Appendix~\ref{Sec:A} contains some useful notation about spinors, and we show explicitly how the enhanced global symmetry emerges.
In Appendix~\ref{Sec:B} we summarize the algebraic properties of $SU(4)$ and $Sp(4)$,
by providing an explicit example of generators for $SU(4)$.
Examples of the renormalized versus bare parameters are given in Appendix~\ref{Sec:C}. 

\section{The model: symmetry considerations}
\label{Sec:model}

The  matter field
content consists of two (massive) Dirac fermions $Q^{i\,a}$, where $a=1,2$ is the $SU(2)$ color index
and $i=1,2$ the flavor index.
The covariant derivative is
 \beqs
 (D_{\mu}Q^i)^a&=&\partial_{\mu}Q^{i\,a}+i g V^A_{\mu}(T^A)^a_{\,\,b}Q^{i\,b}\,,
 \eeqs
 with $V^A_{\mu}$ the gauge fields,  $g$ the coupling, and $T^A$ the generators of $SU(2)$ obeying $\Tr T^AT^B=\frac{1}{2}\delta^{AB}$,
 so that $T^A=\tau^A/2$.
The Lagrangian density is
\beqs
{\cal L}&=&i\,\overline{Q^i}_{\,a}\,\gamma^{\mu}\,(D_{\mu}Q^i)^a\,-\,m\,\overline{Q^i}_{\,a}Q^{i\,a}\,-\,\frac{1}{2}\Tr V_{\mu\nu} V^{\mu\nu}\,,
\eeqs
where the summations over flavor index $i=1,2$ and color index $a=1,2$ are understood, and where the field-strength tensors are defined 
in terms of the gauge bosons 
as $V_{\mu\nu}\equiv \partial_{\mu}V_{\nu}-\partial_{\nu}V_{\mu}+i g \left[V_{\mu}\,,\,V_{\nu}\right]$.
\begin{table}
\begin{center}
\begin{tabular}{|c|c|c|c|}
\hline\hline
{\rm ~~~Fields~~~} &$SU(2)$  &  $SU(4)$\cr
\hline
$V_{\mu}$ & $3$ & $1$ \cr
$q$ & $2$ & $4$ \cr
\hline
$\Sigma_0$ & $1$ & $6$\cr
$M$ & $1$ & $6$\cr
\hline\hline
\end{tabular}
\end{center}
\caption{The field content of the model. $V_{\mu}$ are gauge bosons, $q$ are two-component spinors,
$\Sigma_0$ is a composite scalar, $M$ a scalar spurion.}
\label{Fig:fields}
\end{table}

We collect in Appendix~\ref{Sec:A} and~\ref{Sec:B}  
several useful relations 
between 2-component spinors $q$ and 4-component spinors $Q$, 
as well as details about the algebra of $SU(4)$ and $Sp(4)$
(see also~\cite{sannino,SU4Sp4}).
The global symmetry acting on the matter fields is $U(1)_A\times SU(4)$,
and we explicitly list the transformation properties of the fields in Table~\ref{Fig:fields}.
It is convenient to define: 
\beqs
\Sigma_0^{\,\,nm}&=&\sum_{ab}\,\epsilon_{ab} q^{n\,a\,T} \tilde{C} q^{m\,b}\,,
\eeqs
and to write the mass explicitly as a matrix $M\equiv m\, \Omega$, 
with $\Omega$ the symplectic matrix in Eq.~(\ref{Eq:symplectic}).
The index $n, m=1,\cdots, 4$ and $\tilde{C}=-i\tau^2$ acts on spinor indexes. 
In the lower half of Table~\ref{Fig:fields} we list  the transformation properties of the composite field $\Sigma_0$,
as well as the (symmetry-breaking) spurion $M$.

In the body of the paper, we will describe the finite-temperature properties of composite states that we identify with the pions $\pi$,
$\r$ vector, $a_1$ axial-vector, and $a_0$ scalar mesons.
In the rest of this Section, we summarize the basic properties of  these objects, using the language of effective field theory (EFT).
What results is a Lagrangian density that includes potentially heavy and strongly-coupled degrees of freedom, 
and hence does not yield a calculable weakly-coupled low-energy EFT in the usual sense. 
We use this language to guide our book-keeping exercise, focused on classifying the physical particles, 
their quantum numbers, and the degeneracies---in particular the difference of mass between the $\rho$ and 
$a_1$ vectors and between the  $\pi$ and $a_0$ scalars--- 
that are consequences only of the symmetry structure of the theory and its vacuum.


\subsection{Composite states: scalars}
\label{Sec:scalars}

In the low-energy EFT description, the real antisymmetric field $\Sigma$
 transforms as
\beqs
\Sigma&\rightarrow & U \Sigma U^T
\eeqs
under the action of an element $U$ of $SU(4)$.
The   VEV  $\langle \Sigma \rangle \propto \Omega$
breaks $SU(4)$ to the $Sp(4)$ subgroup.
The generators $T^A$ with $A=1,\cdots,5$ are broken, while  $T^A$ with $A=6,\cdots,15$ are unbroken.
For instance, see \Eq{su4_generators} in Appendix~\ref{Sec:B}.

In terms of the matrix-valued  $\pi(x)=\sum_{A=1}^5 \pi^A(x)T^A$,  the convenient parameterization
\beqs
\Sigma &=&e^{\frac{i\pi}{f}}\Omega e^{\frac{i\pi^T}{f}}\,=\,e^{\frac{2i\pi}{f}}\Omega\,=\,\Omega \,e^{\frac{2i\pi^T}{f}}\,,
\eeqs
automatically satisfies the non-linear constraint $\Sigma^{\dagger}\Sigma=\mathbb{I}_4$. 
The leading-order term of the low-energy EFT is
\beqs
{\cal L}_0&=&\frac{f^2}{4}\Tr\left\{\frac{}{}\partial_{\mu}\Sigma \,(\partial^{\mu}\Sigma)^{\dagger}\right\}\\
&=&\Tr \left\{\frac{}{}\partial_{\mu}\pi\,\partial^{\mu}\pi\right\}\,+\,\frac{1}{3f^2}\Tr\left\{\frac{}{}\left[\partial_{\mu}\pi\,,\,\pi\right]\left[\partial^{\mu}\pi\,,\,\pi\right]\right\}\,+\,\cdots\,.
\eeqs
The pion fields are canonically normalized and hence $f=f_{\pi}$ is the pion decay constant.

The quark mass is incorporated in the EFT by
adding the symmetry-breaking term 
\beqs
{\cal L}_m&=&-\frac{v^3}{4}\Tr \left\{M\,\Sigma\right\}\,+\,{\rm h.c.}
\,=\,2 m v^3 \,-\,\frac{m v^3}{f^2}\Tr \pi^2\,+\,\cdots\,.\label{Eq:Mass}
\eeqs
The expansion in pion fields
 confirms that the $5$ pions are still degenerate, 
 if not massless, in the presence of the explicit breaking given by the Dirac mass for the fermions,
with 
\beqs
m_{\pi}^2f_{\pi}^2&=&{m\, v^3}\,.
\eeqs
The degeneracy of the five pions is a consequence of the unbroken $Sp(4)\sim SO(5)$ symmetry.
The spurion $M$ formally transforms as $M\rightarrow U^{\ast} M U^{\dagger}$, so that if it were promoted to a field
then ${\cal L}_m$ would be manifestly invariant under the full $SU(4)$ symmetry.

Here we pause to make two general observations.
In the context of composite-Higgs models, the presence of a (small) mass term for the quarks is allowed,
contrary to the TC case. While in the latter the quark mass 
explicitly breaks the gauge symmetries, in the composite-Higgs case the SM gauge group is a subgroup of $Sp(4)$,
and hence the term in Eq.~(\ref{Eq:Mass}) does not break it.
The distinction between TC and composite-Higgs cases reduces (in the massless case) to 
a vacuum alignment issue driven by the weak gauging of the $SU(2)_L\times U(1)_Y$ symmetry. In the presence of a mass of the form
in Eq.~(\ref{Eq:Mass}), this problem has a trivial solution: the mass $m$  stabilizes the composite-Higgs vacuum.
Yet, some caution is in order: if $m_{\pi}$ is large, it might become impossible to induce electro-weak symmetry breaking.
We leave these and similar issue out of this study (see~\cite{VA}), 
as in our numerical work all calculations are done with the $SU(2)$ theory in isolation.

To describe the regime in which the symmetry is restored, which is expected to be realized at
high temperature, we remove the non-linear constraint, and hence replace $\Sigma$ by the field $H\sim 6$,
that transforms as a complete 
antisymmetric representation of $SU(4)$.
The kinetic term is
\beqs
{\cal L}_H&=&\frac{1}{2}\Tr \partial_{\mu} H \partial^{\mu} H\,.
\eeqs
The Lagrangian density contains a potential as any arbitrary function $V(H^{\dagger}H)$ is allowed by the symmetries.
The minimization of $V$ yields  the identification
$\langle H\rangle=\frac{f}{\sqrt{2}} \langle \Sigma \rangle=\frac{f}{\sqrt{2}} \Omega$.
The small fluctuations of $H$  are parameterized in terms of the $5$ pion fields
along the broken directions, plus an additional real scalar $\sigma$:
\beqs
H&=&\frac{f+\sigma}{\sqrt{2}}\Sigma\,=\,\frac{f+\sigma}{\sqrt{2}}e^{\frac{2i\pi}{f}}\Omega\,,
\eeqs
where the normalizations are chosen so that all the fields have canonical kinetic terms.
Unconstrained by  symmetry considerations, the scalar
$\sigma$ (singlet of $Sp(4)$) is expected to have a large mass $m_{\sigma}$, 
and in general decay fast to pions.

Besides the $SU(4)\rightarrow Sp(4)$ breaking, the vacuum also induces the  breaking of the 
(anomalous) $U(1)_A$. To discuss it,
we need to promote $H$  a complex field, hence doubling the field content.
We define 
\beqs
\tilde{H}&\equiv&H\,+\,i\,H^{\prime}\,,
\eeqs
with $H^{\prime}$ a second real antisymmetric representation of $SU(4)$.
The action of  $U(1)_A$ is 
\beqs
U(1)_A: && \tilde{H}\,\rightarrow\,e^{i\theta}\tilde{H}\,,
\eeqs
where $\theta$ is the parameter of the $U(1)_A$ transformation.
The  field $H^{\prime}$ introduces an additional $Sp(4)$ singlet
that is the analog of the $\eta^{\prime}$ in QCD and $5$ additional scalars that form a multiplet of
the $SO(5)$ unbroken symmetry, and are the analogue of the $a_0$ isovectors of QCD.
The treatment presented here is indeed a generalization of what done in the context of the linear-sigma-model
description of low-energy QCD~\cite{LSM}.

The presence of the anomaly  produces a large mass 
for $\eta^{\prime}$.
At high temperatures both the fermion condensate and the effect of the anomaly
are suppressed.
Hence, the mass splitting between $a_0$ and $\pi$ provides a measure of the level of breaking of  $U(1)_A$ 
in addition to global $SU(4)$, 
and can be used to look for $SU(4)\times U(1)_A$ thermal restoration. 
Similar arguments hold in the case of QCD (see for example~\cite{SU3a0,SU3eta} and references therein).

Because the $\sigma$ and $\eta^{\prime}$ are flavor singlets, and
the flavor-singlet sector of the spectrum is more difficult to study numerically than the flavored channels,
we will study the $a_0$-$\pi$ mass splitting in order to discuss the restoration of the axial $U(1)_A$ at high temperatures.
We will do so in the body of the paper, using numerical techniques based on the formulation of the theory on
anisotropic lattices.

\subsection{Composite states: vectors}
\label{Sec:vectors}

\begin{figure}
\begin{center}
\includegraphics[width=.35\textwidth]{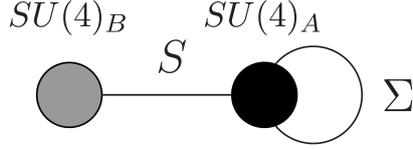}
\end{center}
\caption{The moose diagram representing the EFT description of the vector mesons in the model. }
\label{Fig:EFT}
\end{figure}

The full set of spin-1 vector and axial-vector mesons
spans the adjoint representation of the $SU(4)$ global symmetry. 
A cartoon representing the  EFT description of their long-distance dynamics is 
depicted in Fig.~\ref{Fig:EFT}, and represents a generalization of hidden local symmetry~\cite{HLS,HLS2,HLS3}.
One extends the symmetry from $SU(4)$ to $SU(4)_A\times SU(4)_B$,
with $SU(4)_A$ weakly gauged, with coupling $g_{\r}$. Then one enlarges the field content to include two 
non-linear sigma-model fields $S$ and $\Sigma$.
The non-linear sigma-model $S$ transforms as the bifundamental of $SU(4)_B\times SU(4)_A$, while the field $\Sigma$ transforms on the 
antisymmetric of $SU(4)_A$:
\beqs
S&\rightarrow & U_B\, S \,U_A^{\dagger}\,,~~~~~\Sigma\,\rightarrow\, U_A \Sigma U_A^T\,.
\eeqs
In a composite-Higgs model, the SM gauge group $SU(2)_L\times U(1)_Y$ is a subgroup of $SU(4)_B$.

The gauging of the $SU(4)_A$ symmetry means that (for  global $SU(4)_B$) one has to introduce the covariant derivatives
\beqs
D_{\mu}S &=& \partial_{\mu} S \,-\,i\,g_{\r} S A_{\mu}\,,\\
D_{\mu}\Sigma &=& \partial_{\mu} \Sigma \,+\,i\,g_{\r}\left(A_{\mu}\Sigma\,+\,\Sigma A_{\mu}^T\right)\,,
\eeqs 
and then ${\cal L}_0$  is replaced by all possible 2-derivative invariant operators made by $S$, $\Sigma$, $D S$, $D\Sigma$,
together with the kinetic term for the gauge bosons.
Both $S$ and $\Sigma$ are non-vanishing in the vacuum, inducing the symmetry breaking pattern $SU(4)_A\times SU(4)_B\rightarrow Sp(4)$,
and all vectors are massive.
$\langle \Sigma \rangle$  splits the mass of the $5$ $a_1$  and the $10$ $\r$ mesons.

In unitary gauge, besides the heavy vectors only the
physical pions are retained. They are linear combinations of the fluctuations of $S$ and $\Sigma$. 
The mass term for the pions is
\beqs
{\cal L}_m&=&-\frac{v^3}{4}\Tr\left\{\frac{}{} M\,S\,\Sigma \,S^T \right\}\,+\,{\rm h.c.}\,.
\eeqs
The quark masses also contribute to the masses of the spin-1 states in a more complicated way,
that will be discussed elsewhere~\cite{elsewhere}.

In the absence of the antisymmetric condensate (for $\langle \Sigma \rangle =0$),  $\r$ and $a_1$ mesons would be exactly degenerate.
Their mass splitting is hence a measure of the amount of breaking $SU(4)\rightarrow Sp(4)$.
In the main body of the paper we use the mass splitting between $\r$ (vector) and $a_1$ (axial-vector) as a way to test
whether the  global symmetry is restored at high temperatures.
The generalization to the case in which $\Sigma$ is replaced by $\tilde{H}$ does not require any new ingredients.
In particular the restoration of the axial $U(1)_A$ and of the global $SU(4)$ can, at least in principle, be treated independently.
We summarize in Table~\ref{Fig:particles} the properties of the states discussed in the body of the paper.
One of the purposes of this paper is to make the first steps towards a quantitative assessment of the relation between the 
two phenomena at high temperature, in the specific theory of interest here.

\begin{table}
\begin{center}
\begin{tabular}{|c|c|c|c|}
\hline\hline
${\rm ~~Label~~}$ & ${\rm ~~~Operator~~~}$ & ${\rm~~Meson~~}$  &  $J^{P}$\cr
$S$ & $\overline{Q^i} Q^j$ & $a_0$ & $0^{+}$\cr
$PS$ & $\overline{Q^i}\gamma_5 Q^j$ & $\pi$ & $0^{-}$\cr
$V$ & $\overline{Q^i}\gamma_{\mu} Q^j$ & $\r$ & $1^{-}$\cr
$AV$ & $\overline{Q^i}\gamma_5\gamma_{\mu} Q^j$ & $a_1$ & $1^{+}$\cr
\hline\hline
\end{tabular}
\end{center}
\caption{Interpolating operators, and corresponding flavored particles (i.e. $i\ne j$ in the interpolating operators), 
studied in the body of the paper.
Color and spinor indexes (summed over) are understood.}
\label{Fig:particles}
\end{table}
\section{Numerical results: Anisotropic lattice}
\label{Sec:anisotropic_lattice}
\subsection{Lattice action}
\label{sec:lattice_action}

In this Section, we describe the discretized Euclidean lattice action used for 
our numerical study. For the gauge sector, we modify the standard plaquette action by 
treating the operators containing temporal gauge links separately from those solely containing 
spatial links,
\beq
S_g[U]=\frac{\beta}{\xi^0_g}
\left[
\sum_i (\xi^0_g)^2 \left(1-\frac{1}{N}\textrm{Re}~\textrm{tr}\mathcal{P}_{0i}\right)
+\sum_{i<j}\left(1-\frac{1}{N}\textrm{Re}~\textrm{tr}\mathcal{P}_{ij}\right)
\right],
\label{eq:gauge_action}
\eeq
where $\beta=2N/g^2$ and $\xi^0_g$ are the lattice bare gauge coupling
and the bare gauge anisotropy, respectively. The plaquette $\mathcal{P}$ is defined by
\beq
\mathcal{P}_{\mu\nu}(x)=U_\mu(x)U_\nu(x+\hat{\mu})U_\mu^\dagger(x+\hat{\nu})U_\nu^\dagger(x),
\eeq
where $U_\mu(x)$ denotes the link variables. 
For the fermion sector, we use the Wilson action for fermions in the fundamental represention
\be
S_f[U,\bar{Q},Q]=a_s^3 a_t\sum_x
\bar{Q}(x)D_m Q(x),
\label{eq:fermion_action}
\ee
with the massive Wilson-Dirac operator given by
\be
D_m Q(x)=m_0 Q(x)+\frac{1}{2}\sum_\mu v_\mu
[\gamma_\mu(\nabla_\mu+\nabla^*_\mu)-a_\mu\nabla^*_\mu\nabla_\mu] Q(x),
\label{eq:dirac_eq}
\ee
where $\nabla$ and $\nabla^*$ denote the forward and backward covariant derivatives, respectively:
\bea
\nabla_\mu Q(x)&=&\frac{1}{a_\mu}[U_\mu(x) Q(x+\hat{\mu})-Q(x)],\nn \\
\nabla^*_\mu Q(x)&=&\frac{1}{a_\mu}[Q(x)-U^\dagger_\mu(x-\hat{\mu})Q(x-\hat{\mu})].
\eea
The ratio $v_{\mu}$ of the bare fermion to gauge anisotropy is introduced 
as it can be different to unity. From the redefinition of the fermion field ($Q\rightarrow \sqrt{v_t} Q$ and  $m_0\rightarrow m_0/v_t$),
along with the introduction of the fermion anisotropy $\xi^0_f=\xi^0_g/(v_s/v_t)$, we rewrite
\Eq{dirac_eq} as
\bea
D_m Q(x)&\equiv& (D+m_0) Q(x)\nn 
\,=\, \frac{1}{a_t}\left[\left(a_t m_0+1+\frac{3}{\xi^0_f}\right) Q(x)\right.\nonumber\\
&&\left.\frac{}{}
-\frac{1}{2}\left((1-\gamma_0)U_0(x) Q(x+\hat{0})+
(1+\gamma_0)U_0^\dagger(x-\hat{0}) Q(x-\hat{0})\right)\right.\nn \\
&&\left.-\frac{1}{2\xi^0_f}\sum_j\left((1-\gamma_j)U_j(x) Q(x+\hat{j})+
(1+\gamma_j)U_j^\dagger(x-\hat{j}) Q(x-\hat{j})\right)\right]\,.
\eea
For the rest of this paper we do not explicitly show the lattice spacings for convenience,  
i.e. $a_t=1$, 
except when we need to distinguish the spatial and temporal lattice spacings 
and to discuss the finite temperature. 

The bare anisotropy parameters, $\xi^0_g$ and $\xi^0_f$, are renormalized such that
physical probes at scales well below the cut-off $\sim 1/a$ exhibit Euclidean symmetry, i.e. $\xi_g=\xi_f=\xi$.
For the input quark mass, $M_q$, we parameterize the renormalized parameters $(\xi_g,\xi_f,M_q)$ as
functions of bare parameters $(\xi^0_g,\xi^0_f,m_0)$.
For a small region in the parameter space, 
we assume that the renormalized parameters are linear in the bare parameters.
We further assume that we are in the region of light quark masses, i.e. $M^2_{PS}\sim M_q$,
and arrive at the form~\cite{AnisoL}
\bea
\xi_g(\xi^0_g,\xi^0_f,m_0)=a_0+a_1 \xi^0_g+a_2 \xi^0_f+a_3 m_0,\nn \\
\xi_f(\xi^0_g,\xi^0_f,m_0)=b_0+b_1 \xi^0_g+b_2 \xi^0_f+b_3 m_0,\nn \\
M_{PS}^2(\xi^0_g,\xi^0_f,m_0)=c_0+c_1 \xi^0_g+c_2 \xi^0_f+c_3 m_0.
\label{eq:renormalized_pm}
\eea
For  each set of bare parameters, nonperturbative determinations of $\xi_g$ and $\xi_f$ are carried out 
through the interquark potential and the relativistic meson dispersion relation, respectively, 
which will be discussed in details in the following subsections.

\subsection{Simulation details}
\label{sec:simulations_detail_anisotropic_lattice}

\begin{table}
\begin{center}
\begin{tabular}{ccccccccc}
\hline \hline
~$m_0$~ & ~$\xi^0_g$~ ~& $\xi^0_f$~ & $N_{conf}$ & ~~$M_{PS}$~~ & ~~$M_V$~~ & ~~$\xi_g$~~ & ~~$\xi_f$~~ & ~~$M_{PS}/M_V$~~ \\
\hline
-0.195 & 4.7 & 4.7 & 200 & 0.1659(8) & 0.1823(10)  & 6.19(7) & 6.34(10)  & 0.910(7) \\
-0.195 & 4.9 & 4.7 & 200 & 0.1544(6) & 0.1709(13) & 6.33(8) & 6.33(9)  & 0.903(8) \\
-0.2 & 4.5 & 4.7 & 300 & 0.1616(5) & 0.1784(8)  & 6.03(6) & 6.28(7) & 0.906(5) \\
-0.2 & 4.7 & 4.5 & 300 & 0.1743(5) & 0.1910(7) & 6.07(7) & 6.12(6) & 0.913(4) \\
-0.2 & 4.7 & 4.7 & 200 & 0.1504(6) & 0.1678(10) & 6.13(6) & 6.41(11)  & 0.896(6) \\
-0.2 & 4.9 & 4.7 & 300 & 0.1399(5) & 0.1589(7) & 6.42(6)  & 6.35(7)  & 0.880(5) \\
-0.2 & 5.1 & 4.7 & 160 & 0.1279(13) & 0.1479(19)  & 6.58(9)  & 6.34(17) & 0.865(14)  \\
-0.209 & 4.7 & 4.5 & 150 & 0.1455(7) & 0.1643(11)  & 6.10(6) & 6.04(10) & 0.885(7) \\
-0.209 & 4.7 & 4.7 & 300 & 0.1169(7) & 0.1392(13) & 6.22(6) & 6.35(12)   & 0.840(10) \\
-0.209 & 4.9 & 4.5 & 300 & 0.1336(6) & 0.1533(9) & 6.34(7) & 6.11(9) & 0.872(6) \\
-0.209 & 4.9 & 4.7 & 150 & 0.1023(9) & 0.1243(15) & 6.35(6) & 6.25(12) & 0.823(12) \\
-0.215~\tablefootnote{
This ensemble is used only for the determination of $a_i$ and $c_i$ 
as the number of configurations is not large enough to determine $\xi_f$ in a reliable manner. 
} & 4.7 & 4.7 & 138 & 0.0904(21) & 0.118(5) & 6.04(9) & $\cdot$  & 0.77(3)  \\
-0.209~\tablefootnote{For this ensemble we carry out the measurements on the $128\times 10^3$ lattice. 
Note that $M_{PS}N_s\sim 7$. 
Compared to the $128\times 12^3$ lattice, we find no significant differences in all measured quantities. 
As $M_{PS}N_s\gtrsim 7$ for other ensembles, we therefore expect that 
the finite volume effects are negligible in the tuning of bare lattice parameters. 
} & 4.7 & 4.7 & 300 & 0.1172(7) & 0.1382(11) & 6.13(6) & 6.42(13)  & $\cdot$  \\ \cline{1-9}
\end{tabular}
\end{center}
\caption{%
\label{tab:simulation_details}%
Simulation parameters and results for the tuning of the lattice bare parameters of an anisotropic lattice. 
The masses of pseudoscalar (PS) and vector(V) mesons are measured in units of $a_t$. 
}
\end{table}

We consider the lattice action in \Eq{gauge_action} and \Eq{fermion_action} with two mass-degenerate Wilson fermions.
Configurations are generated using the Hybrid Monte Carlo(HMC) algorithms with the second order Omelyan integrator for Molecular Dynamics(MD) evolution,
where different lengths of MD time steps $\delta \tau_\mu$ are used for gauge and fermion actions such that the acceptance rate is in the range of $75-85\%$.
The simulation codes are developed from the HiRep code~\cite{HiRep} modified by implementing the gauge and fermion 
anisotropies described in Section~\ref{sec:lattice_action}.
To optimise the acceptance rate, 
we also treat the variance of temporal and spatial conjugate momenta differently by introducing a new tunable parameter~\cite{MSU}, 
which is essentially equivalent to the multiscale anisotropic molecular dynamics update~\cite{AnisoL}. 
Without changing the validity of the algorithm, such a setup is helpful for the anisotropic lattice calculations 
through balancing the temporal and spatial MD forces: 
typically the former is larger than the latter approximately by the anisotropy in the lattice spacings.

Except the lattice of $N_t\times N_s^3=128\times 10^3$ for the investigation of finite volume effects, 
all of the numerical calculations for the tuning of bare parameters are performed on 
$N_t\times N_s^3=128\times 12^3$ lattices. We use periodic boundary conditions 
in each direction of both link variables and fermion fields.~\footnote{
We have checked that using antipeoriodic boundary conditions in the time direction for fermions 
give compatible results as expected in zero-temperature calculations.}
Twelve ensembles are created with different bare quark masses, gauge and fermion anisotropies at $\beta=2.0$,
where the details are found in \Table{simulation_details}.
Thermalization and autocorrelation times are estimated by monitoring the average plaquette expectation values.
For each ensemble $N_{conf}=138-300$ configurations are accumulated after $200$ trajectories for thermalization,
where every two adjacent configurations are separated by one auto-correlation length 
of which the typical size is $8\sim 12$ trajectories. 
The statistical errors for all quantities extracted in this work are obtained using the standard bootstrapping technique.

\subsection{Gauge anisotropy}
\label{sec:gauge_anisotropy}

The gauge anisotropy $\xi_g$ is determined from the static potential using Klassen's method~\cite{Klassen}. 
We first define the ratios of spatial-spatial and spatial-temporal Wilson loops by
\be
R_{s}(r,y)=\frac{W_{ss}(r,y)}{W_{ss}(r+1,y)}~\textrm{and}~
R_{t}(r,t)=\frac{W_{st}(r,t)}{W_{st}(r+1,t)},
\label{eq:W_ratios}
\ee
respectively. 
In an asymptotic region, these ratios fall exponentially with the linear interquark potential
and do not depend on $r$, $R_{s}(r,y)\sim e^{-a_s V_s(ya_s)}$ and $R_{t}(r,t)\sim e^{-a_s V_s(ta_t)}$.
Finite volume effects are expected to be suppressed since they are canceled out in the ratios~\cite{Umeda,Klassen}.
As the interquark potential at the same physical distance should yield the same value,
one can extract the anisotropy $\xi_g$ by imposing $R_s(r,y)\equiv R_t(r,t=\xi_g y)$.
In practice, we determine $\xi_g$ by minimizing~\cite{Umeda}
\beq
L(\xi_g)=\sum_{r,y}\ell(\xi_g;r,y),
\label{eq:function_L}
\eeq
with
\beq
\ell(\xi_g;r,y)=\frac{(R_{s}(r,y)-R_{t}(r,\xi_g y))^2}{(\Delta R_s)^2+(\Delta R_t)^2},
\label{eq:function_l}
\eeq
where $\Delta R_s$ and $\Delta R_t$ are the statistical errors of $R_s$ and $R_t$, respectively.

In the original Klassen's approach, the planar Wilson loops are considered where
$r$ is either $x$ or $z$. 
A typical difficulty in this approach is the limited number of data points as 
one quickly encounters a severe signal-to-noise problem in the calculations of the large Wilson loops.
By noting that $\vec{r}$ can be any two-dimensional path in the $x$-$z$ plane with $r=\sqrt{x^2+z^2}$, 
we extend the Klassen's method by including nonplanar Wilson loops
along the closed path $C_y(x,z,y)$ and $C_t(x,z,t)$ with $x\geq z$. 
To maximize the overlap with the physical ground state the shortest paths in the $x$-$z$ plane 
are considered using the Bresenham algorithm which has been applied for the lattice study 
of quark antiquark potential, i.e. see \cite{BA}. 
Analogous to the planar case,
we define $\vec{r}=(x,z)$ and $\vec{r}+1=(x+1,z)$ for a fixed value of $z$.

\begin{figure}
\begin{center}
\includegraphics[width=.85\textwidth]{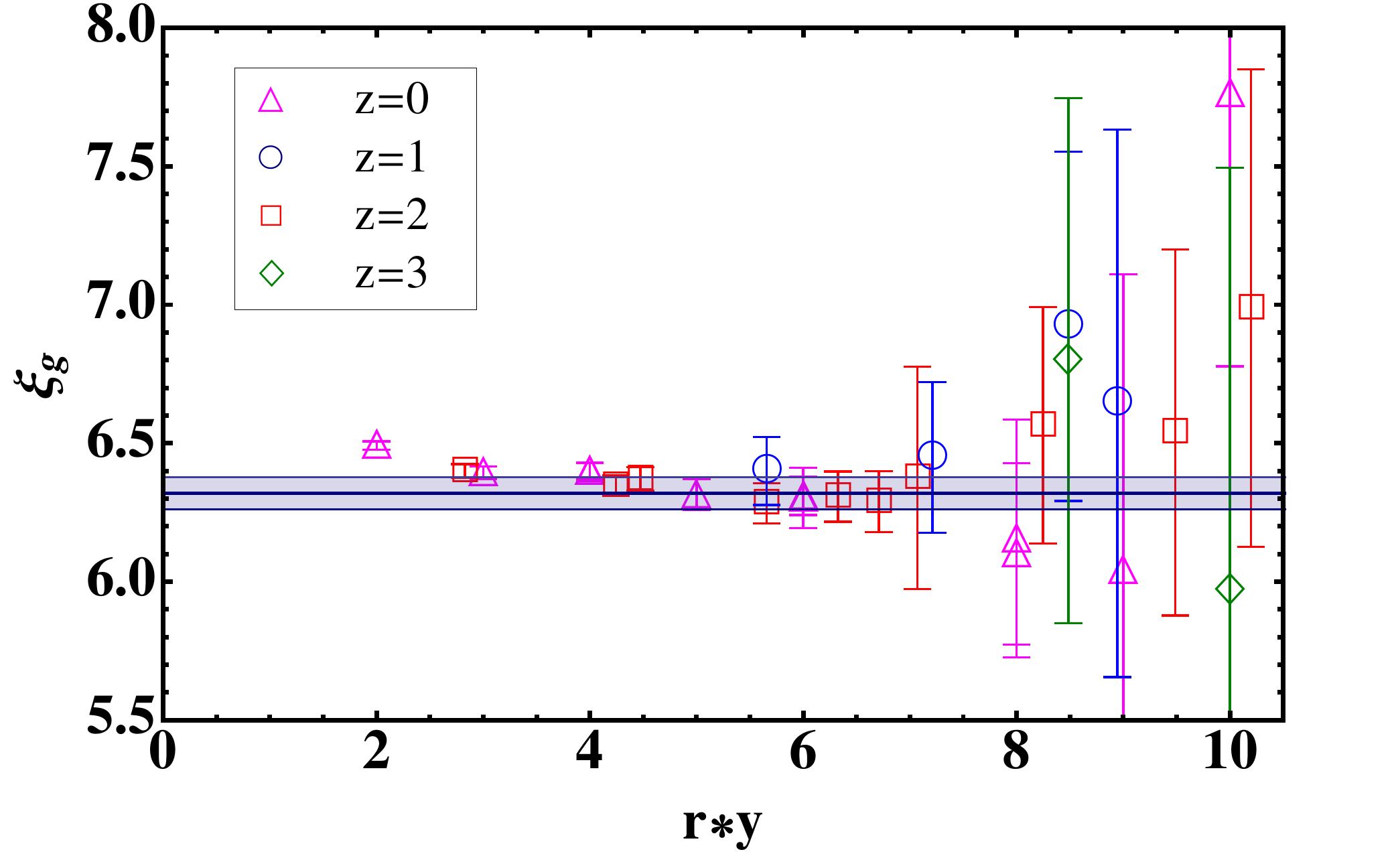}
\caption{%
\label{fig:wilsonloop_ratio_l}%
Lattice artefacts due to nonplanar Wilson loops. 
Different colored points denote the $\xi_g$ obtained by using \Eq{function_l} 
with $z=0,\cdots,3$ for a given $r*y$, while the blue band denotes the extracted value of $\xi_g$. 
The details are found in the main text. 
}
\end{center}
\end{figure}

Using the generalized Klassen's method, we are able to secure enough data points 
having reasonable statistical errors. 
As a consequence, not only do we find the clean signal of an asymptotic region 
in which $\xi_g$ converges,
but also reduce the statistical error of the gauge anisotropy $\xi_g$. 
However, due to the breaking of rotational symmetry on the lattice, 
results obtained mixing on-axis and off-axis loops might be affected by 
a large systematics. 
To investigate this issue, we calculate $\xi_g$ by minimizing the function $\ell(\xi_g;r,y)$ 
with $z=0,\cdots,3$, corresponding to the different shapes of the 2-dimensional paths. 
The results are shown in \Fig{wilsonloop_ratio_l}. 
We find no significant deviations between colored data, 
suggesting that any potential effect of the breaking of rotational 
symmetry cancels in the ratios of Wilson loops in \Eq{W_ratios}. 
For $r*y\geq5$ all data points are statistically consistent with one another. 
The measured value of $\xi_g$ is denoted by the blue band in the figure, where 
its extraction is discussed in the following.

\begin{figure}
\begin{center}
\includegraphics[width=.85\textwidth]{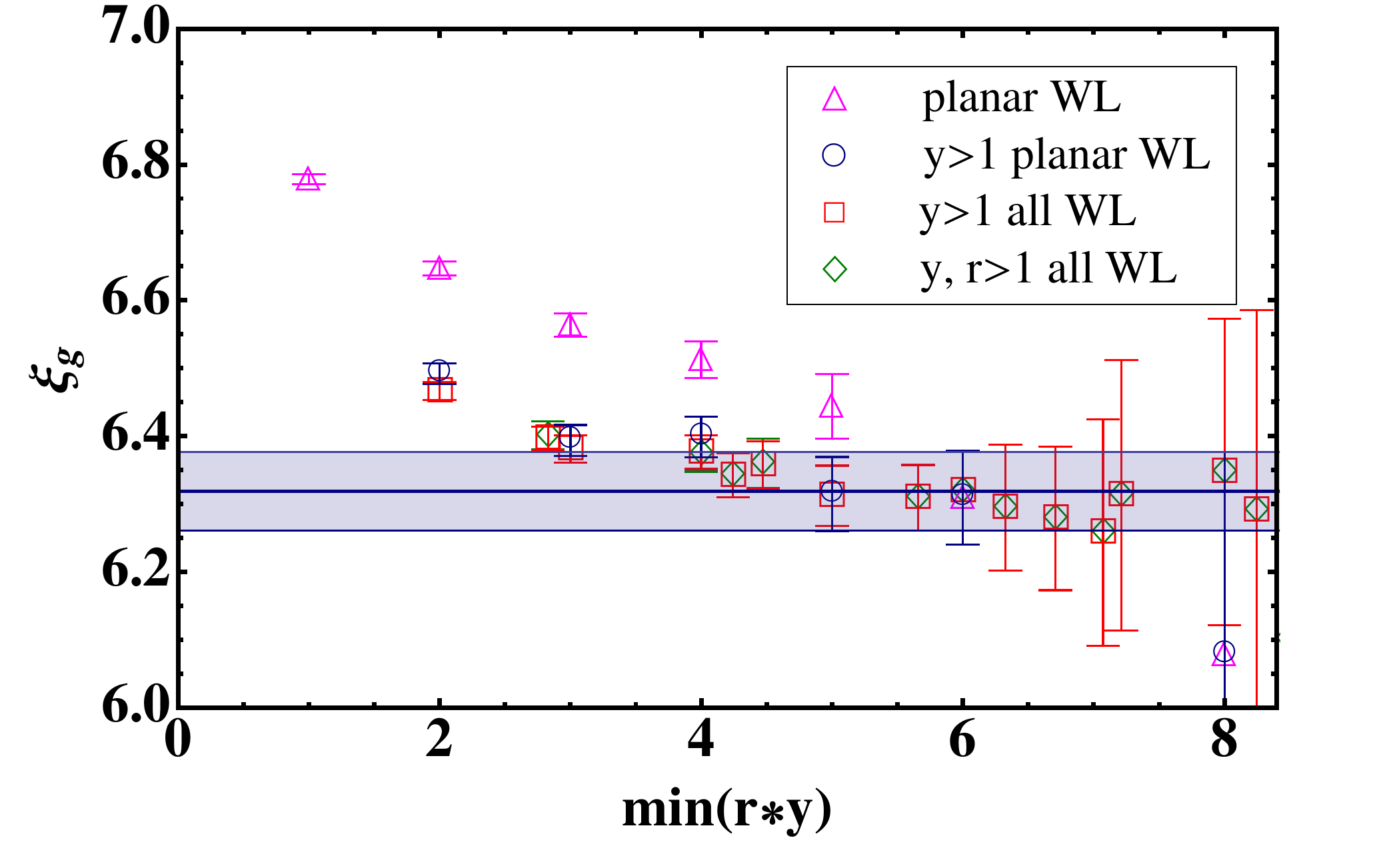}
\caption{%
\label{fig:wilsonloop_ratio}%
Gauge anisotropy $\xi_g$ extracted from the ratios of Wilson loops. 
Different colored points denote the values of  $\xi_g$ obtained by using \Eq{function_L} 
with different sets of data for a given $\textrm{min}(r*y)$, 
while the blue band denotes the extracted value of $\xi_g$. 
Details are found in the main body of the paper.
}
\end{center}
\end{figure}

In \Fig{wilsonloop_ratio} we plot $\xi_g$, obtained by using \Eq{function_L}, 
as a function of $\textrm{min}(r*y)$ 
for four different sets of data:
all planar Wilson loops (purple triangle), planar Wilson loops except $y=1$ (blue circle),
planar and nonplanar Wilson loops except $y=1$ (red square), 
and planar and nonplanar Wilson loops except $y=1$ and $r=1$ (green diamond).
The largest value of $r*y$ is the one before we encounter significant numerical noise. 

For all ensembles we find that $\xi_g$ converges to the asymptotic value
at around $\textrm{min}(r*y)=4\sim 6$ and thus we choose $\textrm{min}(r*y)=6$, 
as for this value we expect the size of systematic errors to be small compared to the statistical error. 
Since the inclusion of $y=1$ Wilson loops causes significant systematic 
effects due to short-range lattice artefacts, as can be seen in the plots 
(see also the discussion in~\cite{Umeda,AnisoL}, in the case of QCD), 
we exclude these Wilson loops for the determination of $\xi_g$. 
In summary, we calculate the asymptotic value of $\xi_g$ 
using planar and nonplanar Wilson loops, except the ones having $y=1$, 
at $\textrm{min}(r*y)=6$ and the results are reported in \Table{simulation_details}.

\subsection{Fermion anisotropy}
\label{sec:fermion_anisotropy}

The fermion anisotropy $\xi_f$ is determined through the leading-order relativistic dispersion relation of mesons
\beq
E^2(p^2)=m^2+\frac{p^2}{\xi_f^2},~~~~\vec{p}=2\pi \vec{n}/N_s,
\label{eq:fermion_dispersion}
\eeq
where $N_s$ is the spatial lattice size. The energy $E$ and the mass $m$ are in units of $a_t$,
while the momentum $\vec{p}$ is in units of $a_s$. 
In the Euclidean formulation, meson two-point correlation functions exponentially fall off 
with the lowest energy at an asymptotically large time. 
In practice, it is useful to define an effective mass, 
\beq
m_{\textrm{eff}}(t)=\cosh^{-1}\left(
\frac{C(t+1)+C(t-1)}{2C(t)}
\right), 
\eeq
where $C(t)$ is the ensemble average of meson correlators. 
Then, ground state energies are obtained from a constant fit to the plateau of 
$m_{\textrm{eff}}$ in the asymptotic region of large $t$. 
In the case of zero momentum these energies are nothing but the meson masses. 
The measured masses of pseudoscalar and vector mesons are reported in \Tab{simulation_details}. 

\begin{figure}
\begin{center}
\includegraphics[width=.85\textwidth]{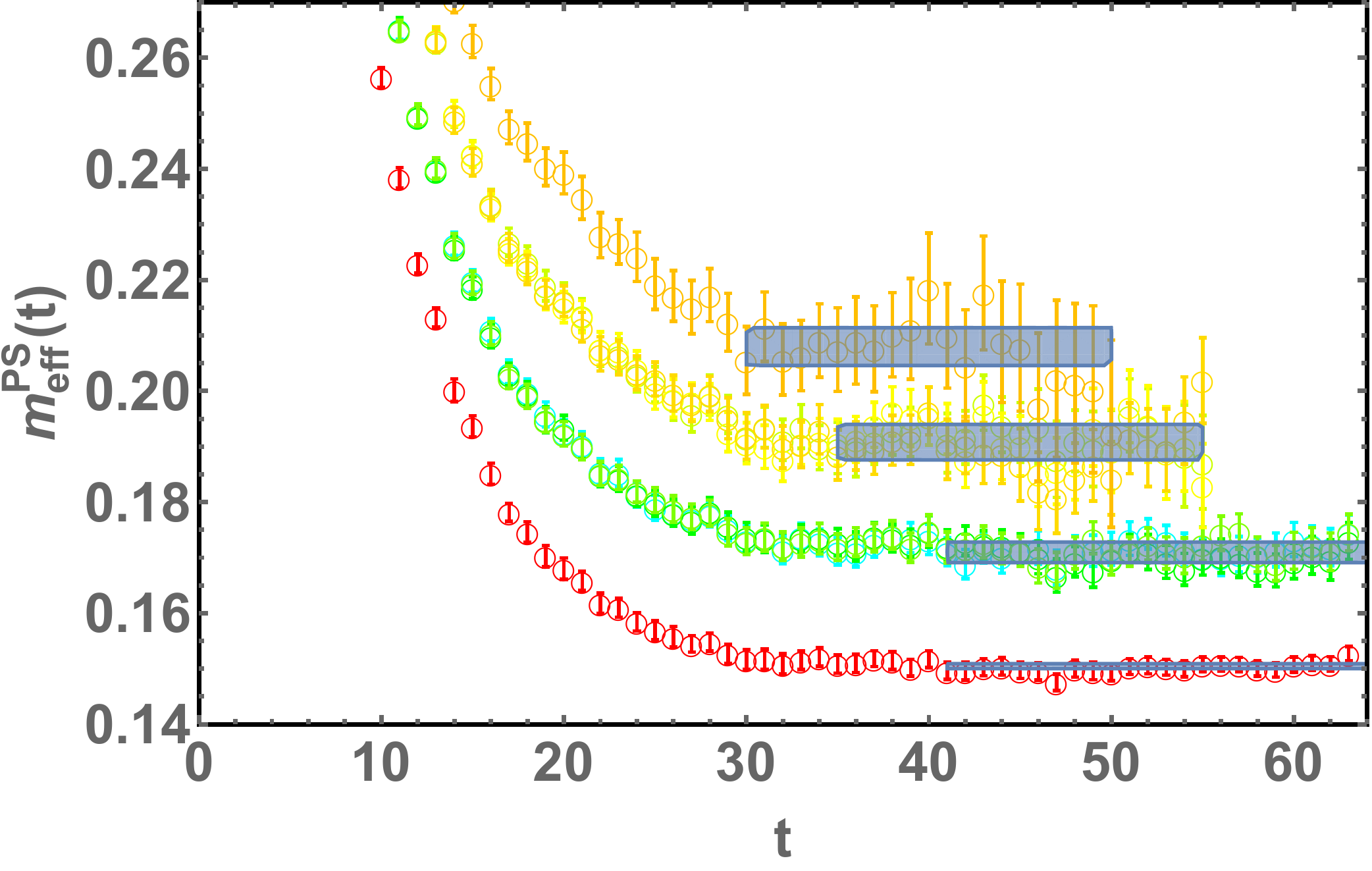}
\includegraphics[width=.85\textwidth]{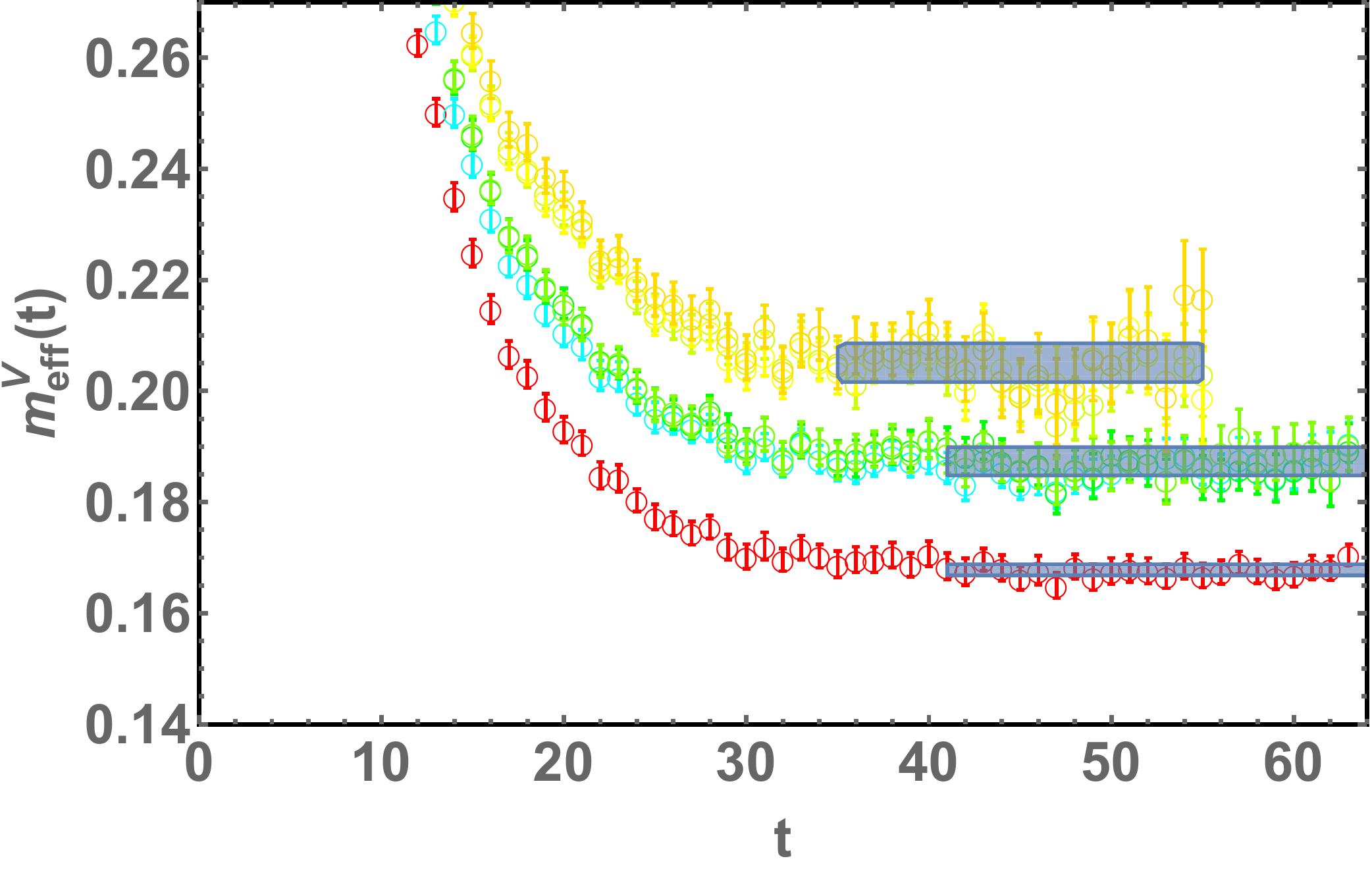}
\caption{%
\label{fig:meff}%
Effective mass plots for a pseudoscalar meson $m_{eff}^{PS}$ and a vector meson $m_{eff}^V$. 
Red, green, yellow, brown colors represent 
different momenta $|\vec{n}|=0,1,2,3$, respectively. 
The blue bands denote the ground state energies obtained by fitting the effective mass to a constant 
in the asymptotic region. 
The lattice bare parameters used in these plots are 
$\beta=2.0$, $m_0=-0.2$, $\xi^0_g=4.7$, $\xi^0_f=4.7$, and $N_t\times N_s^3=128\times 12^3$.
}
\end{center}
\end{figure}

\begin{figure}
\begin{center}
\includegraphics[width=.85\textwidth]{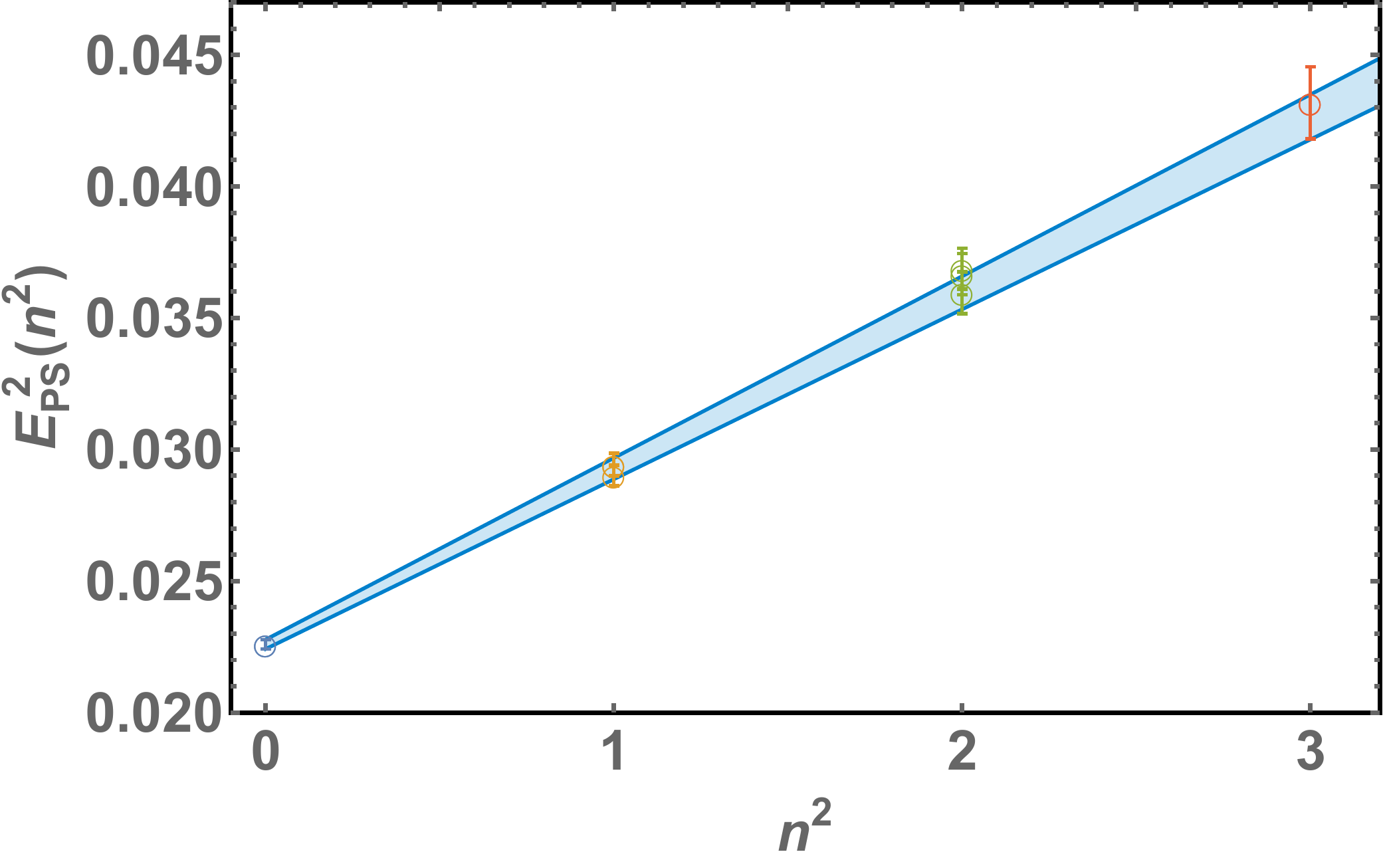}
\includegraphics[width=.85\textwidth]{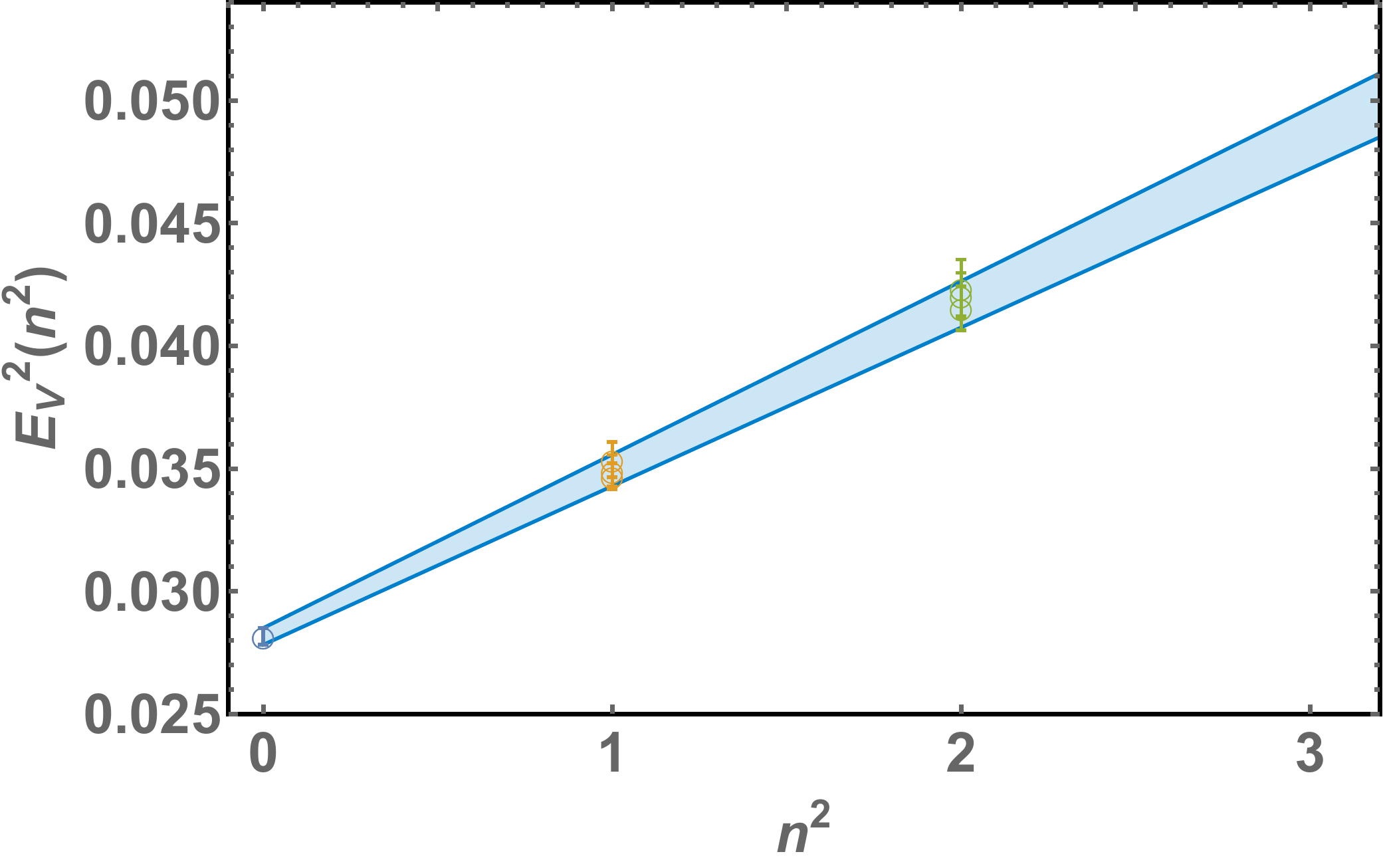}
\caption{%
\label{fig:E2_vs_n}%
Squared energy as a function of squared momentum $|\vec{n}|^2$ for a pseudoscalar meson $E_{PS}^2$ 
and a vector meson $E_V^2$. 
The blue band is obtained by fitting data over $n^2=[0,1]$ to a linear function of \Eq{fermion_dispersion}. 
The lattice bare parameters used in these plots are 
$\beta=2.0$, $m_0=-0.2$, $\xi^0_g=4.7$, $\xi^0_f=4.7$, and $N_t\times N_s^3=128\times 12^3$.
}
\end{center}
\end{figure}

As an example, in \Fig{meff} we show the effective mass plots for pseudoscalar and vector mesons 
with  $m_0=0.2$, $\xi_g^0=\xi_f^0=4.7$ and $\beta=2.0$. 
We construct the meson interpolating operators at source and sink using point sources. 
Various momentum projections with $|\vec{n}|=0,1,2,3$ are denoted by red, green, yellow and brown colors, respectively, 
while the measured ground state energies are denoted by the blue bands. 

In \Fig{E2_vs_n} we plot the resulting squared energy $E^2$ as a function of $|\vec{n}|^2$ and 
find a good linearity, consistent with \Eq{fermion_dispersion}. 
In the determination of $\xi_f$, to minimize the systematic effects due to excited state contamination at higher momenta, 
we only use the lowest four momentum vectors 
$\vec{n}=(0,0,0),~(1,0,0),~(0,1,0),~(0,0,1)$ in the linear fit of $E^2(|\vec{n}|^2)$ to \Eq{fermion_dispersion}. 
As seen in the figures, the fit results denoted by blue bands explain the data very well. 
The extracted value of $\xi_f=6.41(11)$ from a pseudoscalar meson is in good agreement with 
the one from a vector meson, $\xi_f=6.36(14)$, and shows better precision. 
Therefore, for the tuning of lattice bare parameters we use $\xi_f$ from pseudoscalar mesons 
which are summarized in \Table{simulation_details}.

\subsection{Tuning results}
\label{Sec:tuning_results}

To determine the coefficients, $a_i$, $b_i$, and $c_i$, 
we perform the simultaneous $\chi^2$ fit of the numerical data in \Tab{simulation_details} to the functions in \Eq{renormalized_pm}. 
The results are
\bea
a_0=0.6(16),~a_1=0.97(13),~a_2=0.31(23),~a_3=2(4),\nonu \\
b_0=1.8(24),~b_1=0.06(18),~b_2=1.1(3),~b_3=4(7),\nonu \\
c_0=0.475(5),~c_1=-0.0168(4),~c_2=-0.0375(6),~c_3=0.986(11),
\label{eq:fit_results}
\eea
where the values of $\chi^2$ per degrees of freedom are $1.72$, $0.72$, $0.23$, respectively. 
In Appendix~\ref{Sec:C} we show some examples of the results of the fit
in the two-dimensional spaces of the renormalized and bare parameters. 

Our interpretation of the above results requires that we comment on a few important features. 
First of all, renormalized anisotropies are somewhat larger than 
the bare anisotropies, which we interpret as a signal of the fact that
the calculations are performed far from the weak coupling limit. 
Secondly, we find that the coefficients $a_2$ and $b_1$ are small,
in particular, $b_1$ is zero within the statistical errors. 
In the quenched approximation, one would expect 
that the gauge and fermion anisotropies can be determined independently. 
The mild dependences of $\xi_f$ on $\xi_g^0$ and $\xi_g$ on $\xi_f^0$ 
are consistent with the fact that  this part of the numerical study is performed  in the regime of heavy quarks. 
Yet, we note that over the range of considered lattice parameters 
our results show a good linear dependence of the squared mass of a pseudoscalar meson $M_{PS}^2$ on the bare quark mass $m_0$,
which is consistent with our use of  \Eq{renormalized_pm} to extrapolate to the limit of vanishing physical mass for the quarks.

In order to determine the values of the bare parameters at our chosen reference point 
we impose the following renormalization conditions:
\beq
\xi_g(\xi^{0*}_g,\xi^{0*}_f,m^*_0)=\xi_f(\xi^{0*}_g,\xi^{0*}_f,m^*_0)=\xi,~~~
M^2_{PS}(\xi^{0*}_g,\xi^{0*}_f,m^*_0)=m^2_{ps}.
\eeq
Solving \Eq{renormalized_pm} with our target renormalized parameters of $\xi=6.3$ and $m_{ps}^2=0.005$, 
we find 
\beq
\xi_g^{0*}=4.84(8),~\xi_f^{0*}=4.72(12),~m_0^*=-0.2148(37).
\label{eq:bare_pms}
\eeq
We will use these choices for the lattice parameters in measuring the physical properties of the field theory.
Note that $m_0^*$ falls slightly outside the range of masses used in this part of the study (see \Table{simulation_details}),
and hence we expect some (small) residual quark mass and symmetry-breaking effects to be present in our physical simulations. 


\section{Numerical results: Finite temperature}
\label{Sec:finite_temp}
\begin{table}
\begin{center}
\begin{tabular}{cccccccc}
\hline \hline
~$N_z$~ & ~$N_t$~ & ~$T/T_c$ ~& ~$N_{conf}$ ~&~~~~ $N_z$~ & $N_t$ & $T/T_c$ & ~$N_{conf}$~~  \\
\hline 
16 & 16 & 2.44 & 200 & ~~~~24~ & 8 & 4.88 & 200 \\
& 20 & 1.95 & 200 &  & 12 & 3.25 & 200 \\
& 24 & 1.63 & 200 &  & 16 & 2.44 & 225 \\
& 28 & 1.39 & 200 &  & 20 & 1.95 & 150 \\
& 30 & 1.30 & 200 &  & 24 & 1.63 & 200 \\
& 36 & 1.08 & 200 &  & 28 & 1.39 & 250 \\
& 40 & 0.98 & 200 &  & 36 & 1.08 & 380 \\
& 128 & 0.30 & 215 & & 42 & 0.93 & 388 \\
&  &  &  &  & 48 & 0.81 & 390 \\ 
&  &  &  &  & 56 & 0.70 & 337 \\ \cline{1-8}
\end{tabular}
\end{center}
\caption{%
\label{tab:finite_T_ensembles}%
Details of the ensembles of the $N_t \times 16^3$ and $N_t\times 16^2\times 24$ lattices. 
All calculations use $\xi_g^{0*}=4.84$, $\xi_f^{0*}=4.72$, $m_0^*=-0.2148$ and $\beta=2.0$.
}
\end{table}

\begin{table}
\begin{center}
\begin{tabular}{ccccccc}
\hline \hline
~$T/T_c$ ~& $M_{PS}^S$~ & $M_{S}^S$ & $M_V^S$ & ~~$M_{AV}^S$~~ & ~~$R_S$~~ & ~~$R_V$~ \\
\hline
2.44 & 0.3322(2) & 0.3823(8) & 0.3475(2)  & 0.3879(10) & 0.070(1)  & 0.0549(13) \\
1.95 & 0.2815(6) & 0.3292(14) & 0.3045(5)  & 0.340(2) & 0.078(2)  & 0.055(3) \\
1.63 & 0.2355(6) & 0.281(2) & 0.2629(7)  & 0.2938(18) & 0.088(4)  & 0.056(3) \\
1.39 & 0.1937(13) & 0.236(5) & 0.2272(12)  & 0.248(4) & 0.099(10)  & 0.045(8) \\
1.30 & 0.1783(13) & 0.228(5) & 0.2110(13)  & 0.235(5) & 0.123(11)  & 0.054(10) \\
1.08 & 0.1312(11) & 0.201(6) & 0.1699(15)  & 0.191(5) & 0.210(13)  & 0.057(13) \\
0.98 & 0.1147(12) & 0.180(4) & 0.1525(16)  & 0.185(6) & 0.222(12)  & 0.096(16) \\
0.30 & 0.0758(3) & 0.200(8) & 0.1068(11) & 0.213(9) & 0.449(16)  & 0.331(19)  \\ \cline{1-7}
\end{tabular}
\end{center}
\caption{%
\label{tab:finite_T_lattice_1}%
Simulation results for the $N_t \times 16^3$ lattice. 
To compensate the anisotropy of the Euclidean lattice, 
meson screening masses $M^S$ are obtained from the measured masses divided by $\xi=6.3$. 
All calculations use $\xi_g^{0*}=4.84$, $\xi_f^{0*}=4.72$, $m_0^*=-0.2148$ and $\beta=2.0$.
}
\end{table}

\begin{table}
\begin{center}
\begin{tabular}{ccccccc}
\hline \hline
~$T/T_c$ ~& $M_{PS}^S$~ & $M_{S}^S$ & $M_V^S$ & ~~$M_{AV}^S$~~ & ~~$R_S$~~ & ~~$R_V$~ \\
\hline
4.88 & 0.4724(2) & 0.5021(8) & 0.4770(2)  & 0.5044(7) & 0.0305(9)  & 0.0279(8) \\
3.25 & 0.3885(2) & 0.4293(15) & 0.3975(3)  & 0.432(2) & 0.0498(18)  & 0.041(2) \\
2.44 & 0.33257(17) & 0.3833(8) & 0.34760(18)  & 0.3905(5) & 0.0709(11)  & 0.0581(7) \\
1.95 & 0.2813(5) & 0.3268(15) & 0.3043(4)  & 0.3371(14) & 0.075(3)  & 0.051(2) \\
1.63 & 0.2326(7) & 0.275(2) & 0.2617(7)  & 0.290(3) & 0.083(4)  & 0.052(5) \\
1.39 & 0.1909(9) & 0.234(4) & 0.2239(9)  & 0.251(3) & 0.102(10)  & 0.058(6) \\
1.08 & 0.1295(9) & 0.202(7) & 0.1680(11)  & 0.194(4) & 0.218(17)  & 0.072(11) \\
0.93 & 0.1036(6) & 0.181(4) & 0.1440(11)  & 0.170(5) & 0.272(10)  & 0.081(13) \\
0.81 & 0.0880(5) & 0.191(6) & 0.1254(9)  & 0.186(5) & 0.369(15)  & 0.193(12) \\
0.70 & 0.0798(4) & 0.183(7) & 0.1107(13) & 0.199(6) & 0.393(17)  & 0.284(16)  \\ \cline{1-7}
\end{tabular}
\end{center}
\caption{%
\label{tab:finite_T_lattice_2}%
Simulation results for a $N_t \times 16^2\times 24$ lattice. 
To compensate the anisotropy of the Euclidean lattice, 
meson screening masses $M^S$ are obtained from the measured masses divided by $\xi=6.3$. 
All calculations use $\xi_g^{0*}=4.84$, $\xi_f^{0*}=4.72$, $m_0^*=-0.2148$ and $\beta=2.0$.
}
\end{table}

From now on, the lattice bare parameters are fixed by the central values in \Eq{bare_pms} along with $\beta=2.0$.
We perform finite temperature calculations on the anisotropic lattices 
of $N_t\times16^3$ and $N_t\times16^2\times 24$. 
Simulation details and numerical results for these two lattices are summarized in 
Tables~\ref{tab:finite_T_ensembles}, \ref{tab:finite_T_lattice_1} and \ref{tab:finite_T_lattice_2}.
Two different values of $N_z$ are considered 
to estimate the systematic errors due to excited state contaminations in the calculations of screening masses. 
The algorithms for the generation of gauge ensembles have been discussed in \Sec{simulations_detail_anisotropic_lattice}. 

Before we discuss the numerical results of finite temperature calculations in details, 
we perform a zero temperature calculation in order to check how well the tuned bare parameters are working. 
Using the ensemble of $128\times 16^3$ in \Tab{finite_T_lattice_1}, we obtain 
$\xi_g=6.29(4)$, $\xi_f=6.1(2)$, and $M_{ps}^2=0.00517(14)$. 
These results are compatible with the renormalized parameters of $\xi=6.3$ and 
$m_{ps}^2=0.005$, 
where the largest uncertainty occurs in the detemination of $\xi_f$ with $\sim 3\%$. 
Finite volume effects are expected to be negligible as the lattice volume is much larger than the size of 
the pseudo-scalar meson, $m_{ps} L \sim 7$. 

Adopting anti-periodic boundary condition along the temporal direction, 
temperature is defined by $T\equiv\frac{1}{N_t a_t}$. 
We will find it convenient to
measure the temperature in units of the (pseudo-)critical temperature $T_c$, 
 discussed and measured in the next section.

\subsection{Deconfinement crossover}
\label{sec:deconfinement_crossover}

As is the case for QCD with small number of quarks, 
our model is also expected to exhibit confinement at low temperature and 
form a quark-gluon plasma across the (pseudo-)critical temperature $T_c$. 
Although the Polyakov loop is not an exact order parameter 
when the number of quarks is finite, 
it is widely used as an indicator of deconfinement. 
Following the method used in~\cite{RPM1,RPM2}, 
we define the expectation value of the renormalized Polyakov loop\footnote{
For the discussion of the renormalized Polaykov loop and its scheme dependence, 
we refer the reader to~\cite{RPM3,RPM4}. 
} by
\beq
L_{R}(T)=Z_L^{N_t} L_0(T),
\eeq
where the bare Polyakov loop $L_0(T)$ is related to the bare free energy $F_0(T)$ as 
$L_0(T)=\textrm{exp}(-F_0(T))$. The multiplicative renormaliztion constant is defined by 
$Z_L=\textrm{exp}(-\Delta F_0)$, which only captures the short distant physics and thus is 
independent on the temperature. As different choices of $Z_L$ denote 
different renormalization schemes, to incorporate the scheme dependence on the detemination of $T_c$ 
we impose a renormalization condition for a given temperature $T_R$ by
$L_R(T_R)\equiv \textrm{constant}$. 

\begin{figure}
\begin{center}
\includegraphics[width=.85\textwidth]{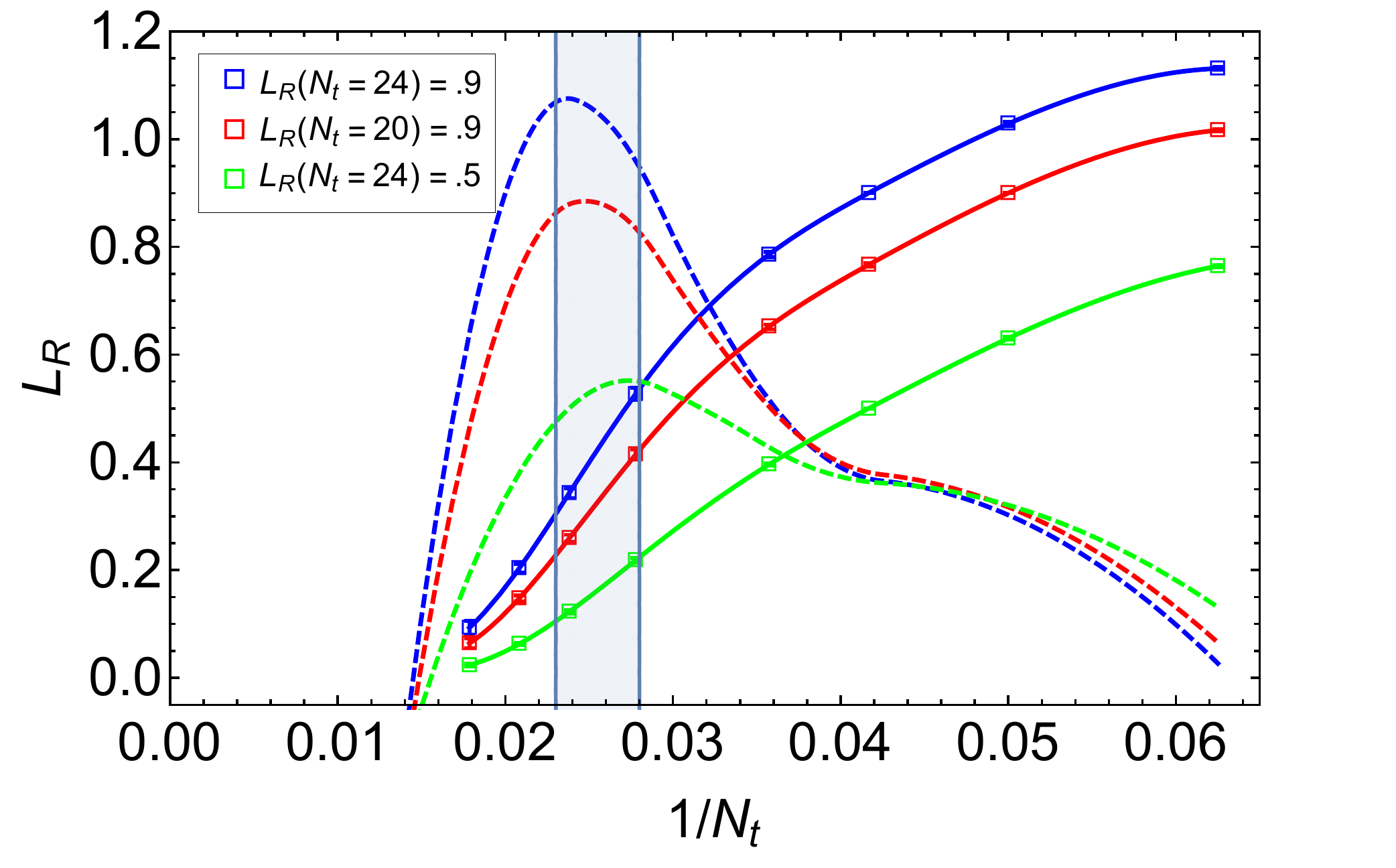}
\caption{%
\label{fig:polyakov_loop}%
Renormalized Polyakov loop and their susceptibility. The renormalized Polyakov loops $L_R$ denoted by 
empty squares are obtained from the ensembles of $N_t\times 16^2\times 24$ with $N_t$ ranged over $[16,56]$. 
The solid curves are the interpolation of $L_R$ connected by cubic splines, while 
the dashed curves are the corresponding susceptibility $\chi(L_R)$, the derivatives of $L_R$ with $N_t^{-1}$. 
Different colors are associated with different renormalization conditions, 
while the blue band denotes the (pseudo-)critical temperature $T_c$ with uncertainties as described in the text. 
}
\end{center}
\end{figure}

We consider three renormalization schemes, defined by the conditions 
$L_R(N_t=24)=0.9$, $L_R(N_t=24)=0.5$, and 
$L_R(N_t=20)=0.9$ respectively. The results are shown in \Fig{polyakov_loop}. 
The temperature $T_c$ is determined from the peak of the susceptibility of the Polyakov loop, 
$\chi(L_R)=\partial L_R/\partial T$, denoted by dashed lines in the figure. 
Combining the statistical uncertainty and the systematic uncertainty of scheme dependences in quadrature, 
we find that $T_c a_t=0.0255(25)$, or equivalently that $N_t^c=39(4)$. 
As anticipated, we will measure temperatures in units of this $T_c$ in the following. 

\subsection{Temporal correlation functions}
\label{sec:temporal_corr}

\begin{figure}
\begin{center}
\includegraphics[width=.85\textwidth]{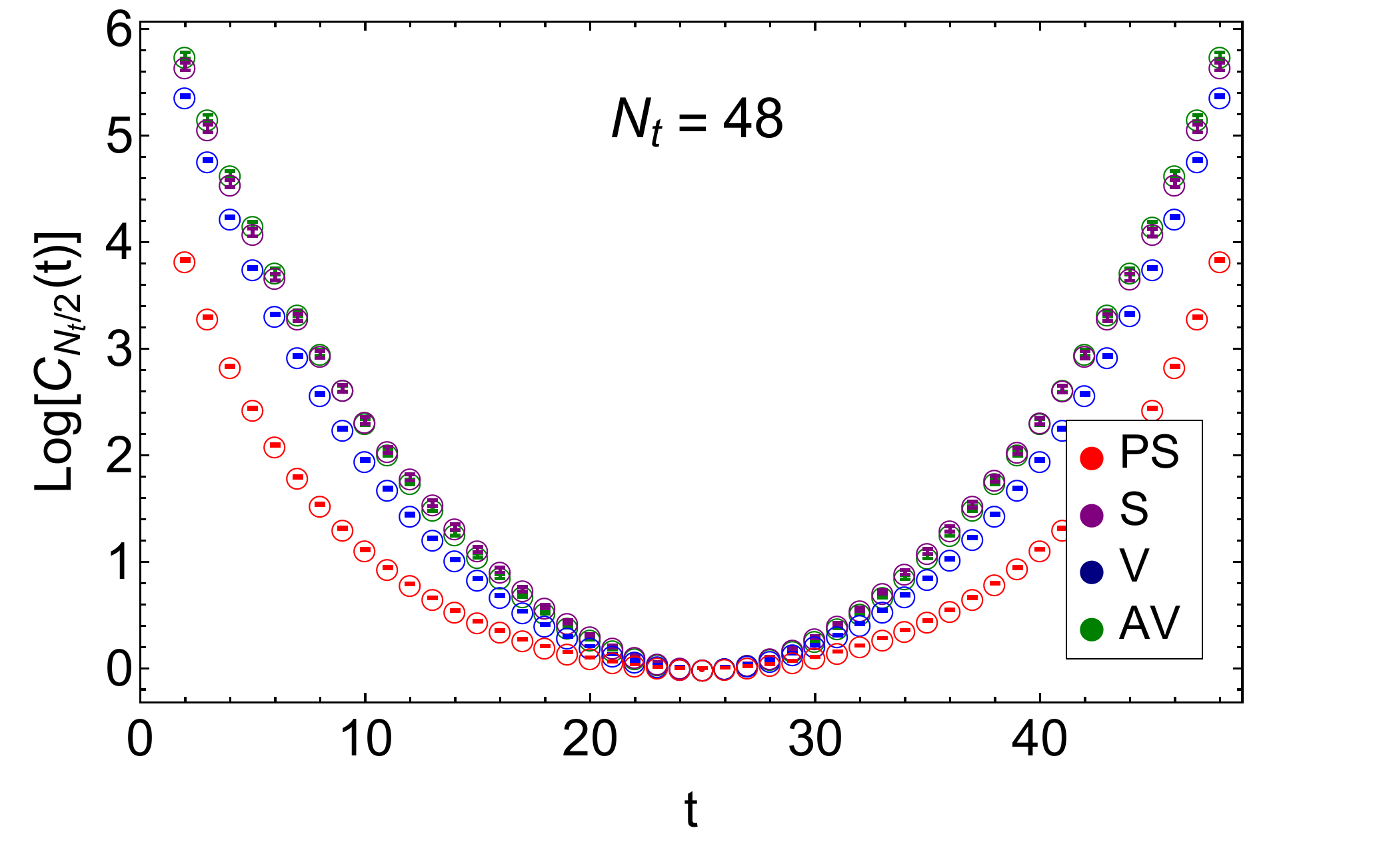}
\includegraphics[width=.85\textwidth]{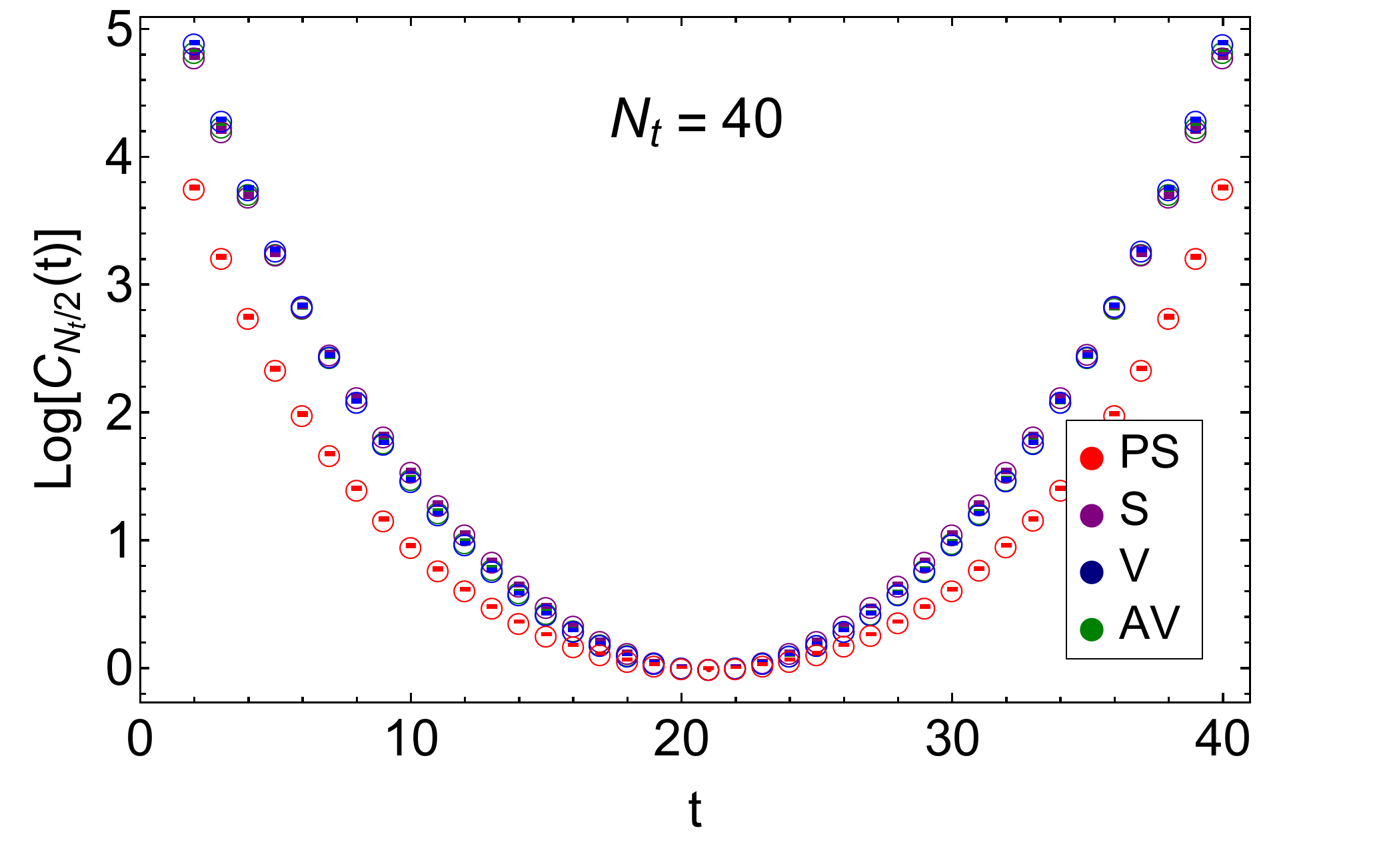}
\caption{%
\label{fig:temp_corr}%
Temporal correlation functions for pseudoscalar(red), scalar(purple), vector(blue), and axial vector(green) mesons. 
For a given Euclidean time $t$, we plot the logarithms of the correlation functions normalized by 
the correlation functions at $N_t/2$. 
}
\end{center}
\end{figure}

At zero temperature, the Euclidean two-point correlation functions of mesonic observables fall off 
with a single exponential at a large time so that the ground state energy of mesons 
can be extracted in a clear way in principle. 
In the finite temperature lattice calculations 
this process is affected by some limitations. 
Firstly, the maximum available physical temporal extent is limited by the inverse of the temperature. 
In addition, a single exponential analysis becomes subtle 
as the spectral function of mesons no longer exhibits a sharp peak at the mass of mesons. 
In this case, it is more desirable to investigate the correlation functions by themselves. 

We introduce the normalized correlation function with the reference choice  $t=N_t/2$: 
\beq
C_{N_t/2}(t)=\frac{ C(t)}{ C(N_t/2)}.
\eeq
We consider isovector pseudo-scalar, scalar, vector, and axial-vector mesons, 
where the corresponding interpolating fields are defined by
\bea
&\mathcal{O}_{PS}(x)=\bar{Q}(x)\gamma_5 Q(x),~
\mathcal{O}_{S}(x)=\bar{Q}(x)Q(x),&\nn \\
&\mathcal{O}_{V}^i(x)=\bar{Q}(x)\gamma^i Q(x),~
\mathcal{O}_{AV}^i(x)=\bar{Q}(x)\gamma_5\gamma^i Q(x),&
\label{eq:meson_ops}
\eea
respectively (flavour indices selecting non-singlet states are understood). 
In order to improve the statistics, we use stochastic wall sources~\cite{SWS} for the study 
of meson spectrum at finite temperature. 
Using these mesonic operators we compute the function $C_{N_t/2}(t)$. 
In \Fig{temp_corr} we show the results of $\log C_{N_t/2}(t)$ for $N_t=48$ and $40$, 
which exemplify the typical behaviors of $C_{N_t/2}(t)$ below and near $T_c$ respectively. 

By comparing the two plots in  \Fig{temp_corr} one can see that while 
at low temperature ($N_t=48$) the vector and axial-vector correlators are 
different, they become hard to distinguish from one another in proximity of $T_c$ ($N_t=40$).
The overlap of $C_{N_t/2}(t)$ between vector and axial-vector mesons 
can be considered as an indication of the parity doubling in the vector channel and thus 
the restoration of the global $SU(4)$ symmetry. 
By contrast, the situation for scalar and pseudo-scalar correlators is quite different, as we will discuss better by
looking at spatial correlation functions in the next subsection, and indicates that at this temperature we do not
yet see evidence of  the restoration of the $U(1)_A$ symmetry.
Notice that the correlation functions still satisfy the Weingarten's mass inequalities~\cite{Weingarten}. 

\subsection{Spatial correlation functions}
\label{sec:spatial_corr}

In contrast to the temporal correlation function, the spatial correlation function at finite temperature 
exhibits a single exponential decay at large time. 
The decay rate is called {\it screening mass}, as it defines the effective length scale 
associated with the excitation of mesonic operators in the medium~\cite{DeTar}. 
At zero temperature the screening mass is equivalent to the meson mass, 
as the temporal and spatial correlation functions share the same spectral function. 

By using the meson interpolating fields in \Eq{meson_ops}, we calculate the ensemble average of spatial correlators $C(z)$
along the $z$-direction  and extract the masses in units of $a_s$ using the analysis method described in \Sec{fermion_anisotropy}. 
Notice that in our anisotropic lattice calculations the spatial and temporal lengths are measured differently. 
To have the consistent lattice unit of mass in $a_t$, we therefore define the screening mass $M^S$ by multiplying 
$\xi^{-1}$ to the measured spatial masses.
In addition to the screening masses, we define the following normalized mass ratios
\beq
R_V(T)=\frac{M_{AV}^S(T)-M_V^S(T)}{M_{AV}^S(T)+M_V^S(T)}\,,
\label{eq:mass_ratio_1}
\eeq
for the vector channel, and 
\beq
R_S(T)=\frac{M_{S}^S(T)-M_{PS}^S(T)}{M_{S}^S(T)+M_{PS}^S(T)}\,,
\label{eq:mass_ratio_2}
\eeq
for the scalar channel.
These quantities are useful to quantify the level of parity doubling in the mass spectrum. 

\begin{figure}
\begin{center}
\includegraphics[width=.85\textwidth]{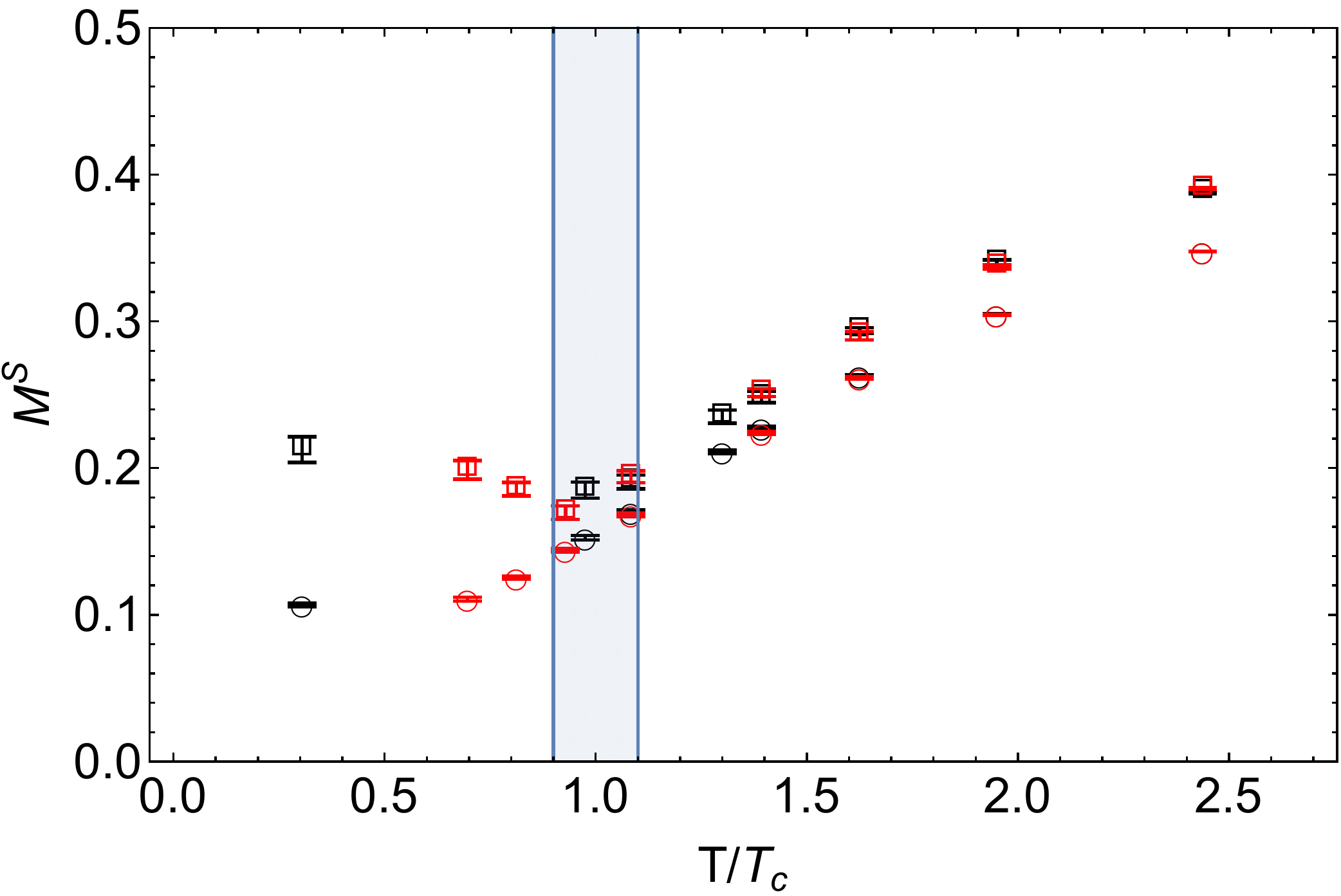}
\includegraphics[width=.85\textwidth]{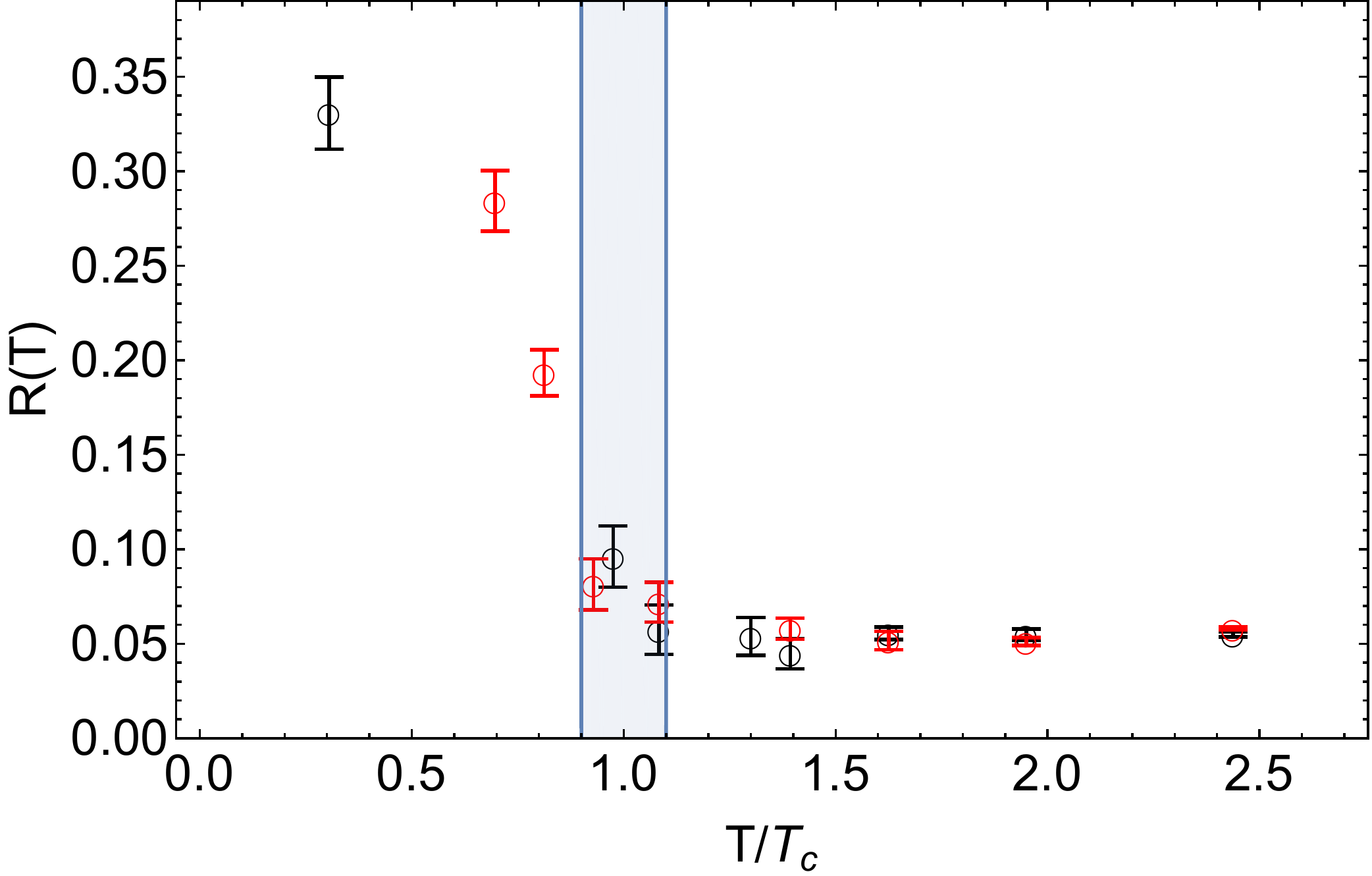}
\caption{%
\label{fig:smass_vector}%
Screening masses of vectors (circles) and axial vector (squares) mesons (top panel) and 
the corresponding mass ratio defined in \Eq{mass_ratio_1} (bottom panel). 
Black and red points are extracted from ensembles with $N_t\times16^3$ and $N_t\times 16^2\times24$ lattice points, respectively. 
The temperature is expressed in units of $T_c$. 
The blue vertical band denotes the uncertainty of $T_c$ itself. 
}
\end{center}
\end{figure}

\begin{figure}
\begin{center}
\includegraphics[width=.85\textwidth]{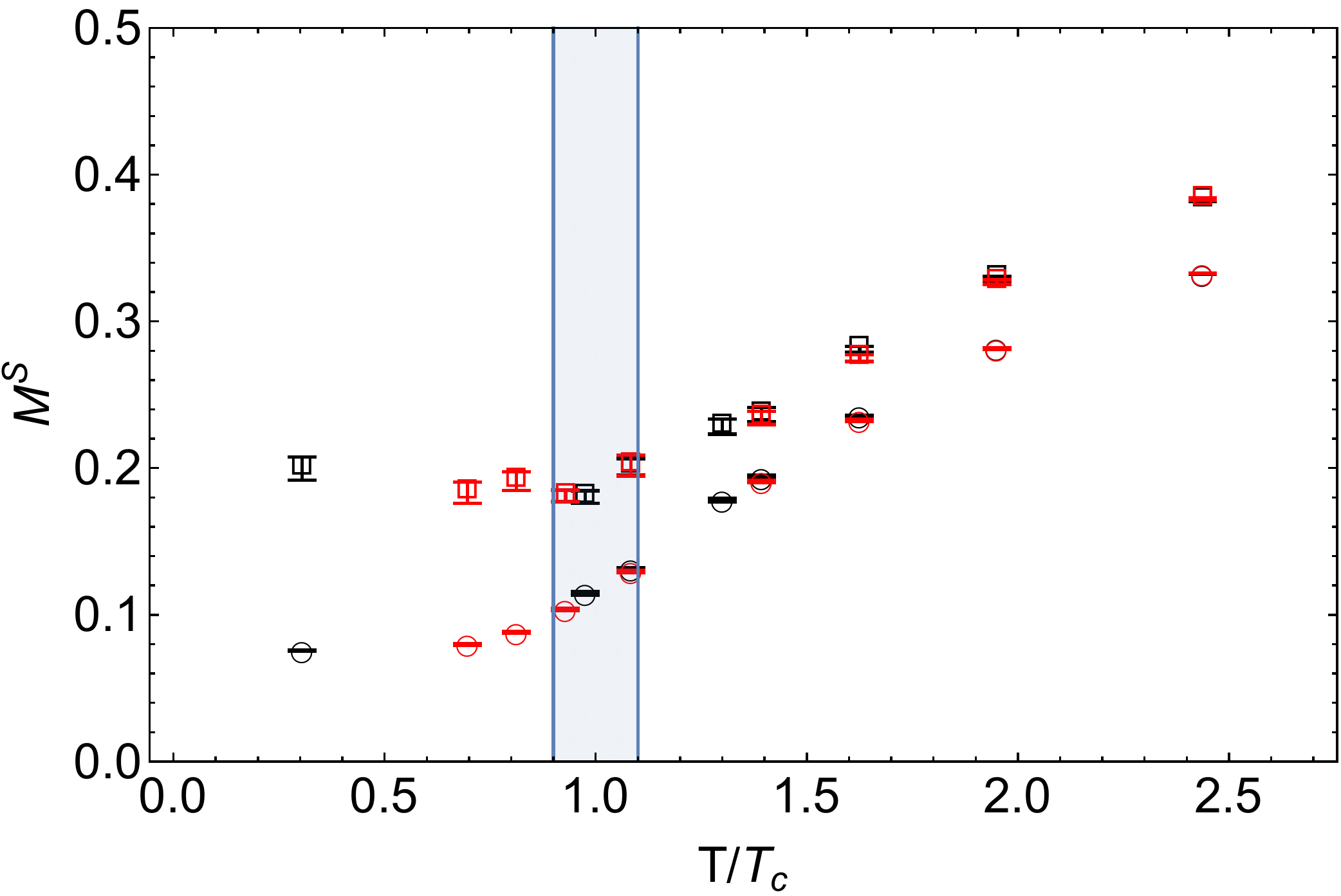}
\includegraphics[width=.85\textwidth]{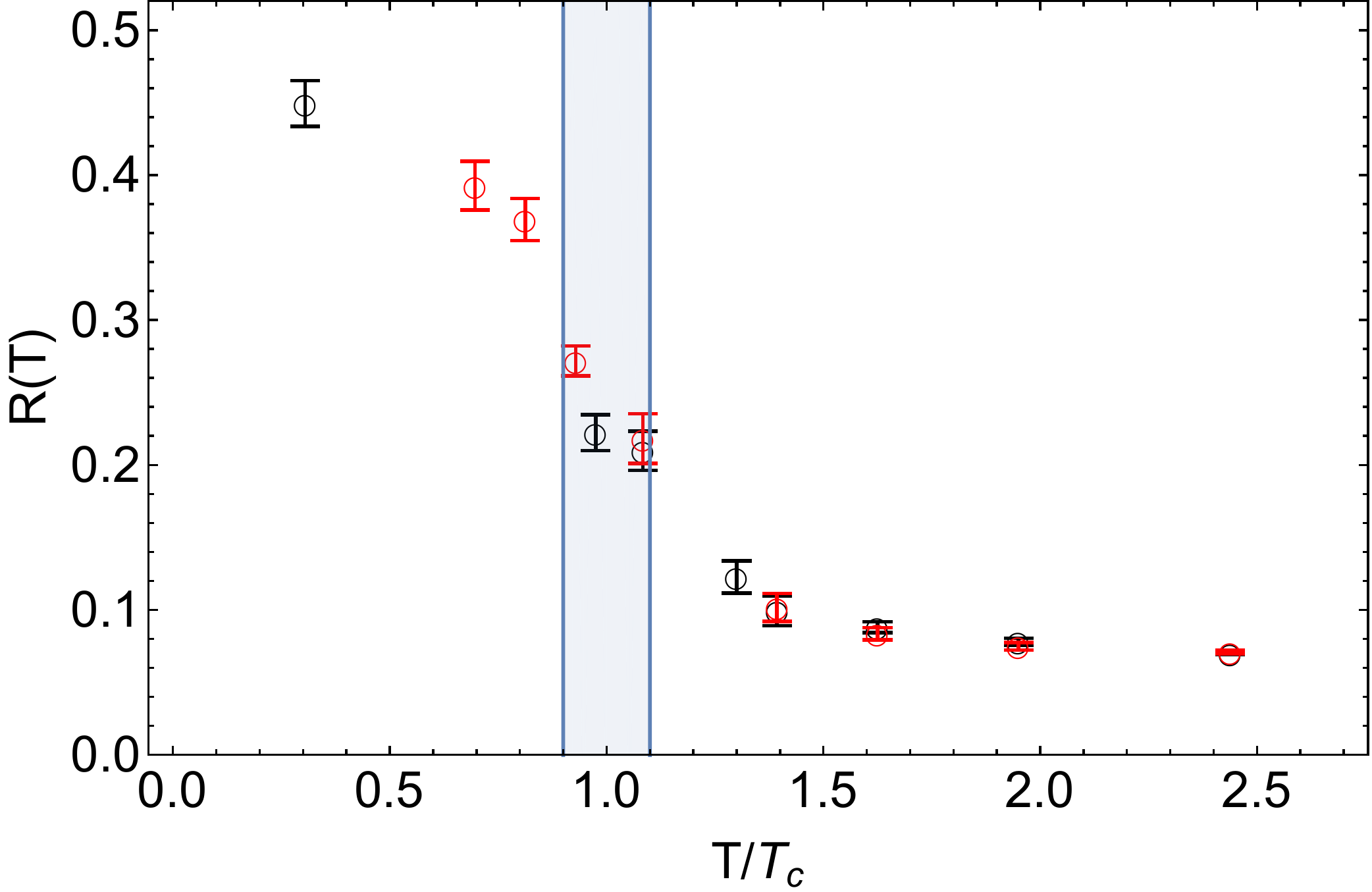}
\caption{%
\label{fig:smass_scalar}%
Screening masses of pseudoscalar (circles) and scalar (squares) mesons (top panel) and 
the corresponding mass ratio defined in \Eq{mass_ratio_2} (bottom panel). 
Black and red points are extracted from ensembles with $N_t\times16^3$ and $N_t\times 16^2\times24$ lattice points, respectively. 
The temperature is expressed in units of  $T_c$. The blue vertical band denotes the uncertainty of $T_c$ itself. 
}
\end{center}
\end{figure}

Our main results are presented in \Tab{finite_T_lattice_1} and~\ref{tab:finite_T_lattice_2}, as well as in 
\Fig{smass_vector} and \ref{fig:smass_scalar}. 
The error bar of each data point only represents the statistical uncertainty. 
We show explicitly the comparison between $N_t\times 16^3$ (black) and $N_t\times 16^2\times24$ (red) lattices.
The level of agreement of the two ensembles  
implies that  there is no significant systematic uncertainty due to excited state contaminations.

By looking first at the vector and axial-vector masses, we see  a plateau in $R_V$ above $T_c$,
which together with the change of behavior of the masses  above $T_c$
strongly suggests that parity partners are degenerate and the global symmetry is effectively restored. 
There is small deviation from zero in the mass ratio at asymptotically large values of $T$, that may be 
the result of finite spacing, finite mass and possibly other small lattice artefacts.

In the case of the scalar channel, the plateau in $R_S$ appears at somewhat larger temperature, $\sim 1.5 T_c$. 
This result may imply that the axial $U(1)_A$ and global $SU(4)$ symmetries are restored 
at different temperatures. However, this is not conclusive, for several reasons.
First of all, because we do not know what kind of transition is appearing in the underlying dynamical model,
and it is likely that $T_c$ actually identifies a cross-over. 
But also because we do not know how much each of the lattice artefacts affects the results,
and it might be that different observables are affected in different amounts by the finite quark mass, 
or the finite value of the coupling.
A relevant discussion in the context of two-flavor QCD can be found in~\cite{SU3a0}, 
for instance, 
where the numerical results strongly suggest that the symmetry restorations occur simultaneously in the massless limit.

\begin{figure}
\begin{center}
\includegraphics[width=.85\textwidth]{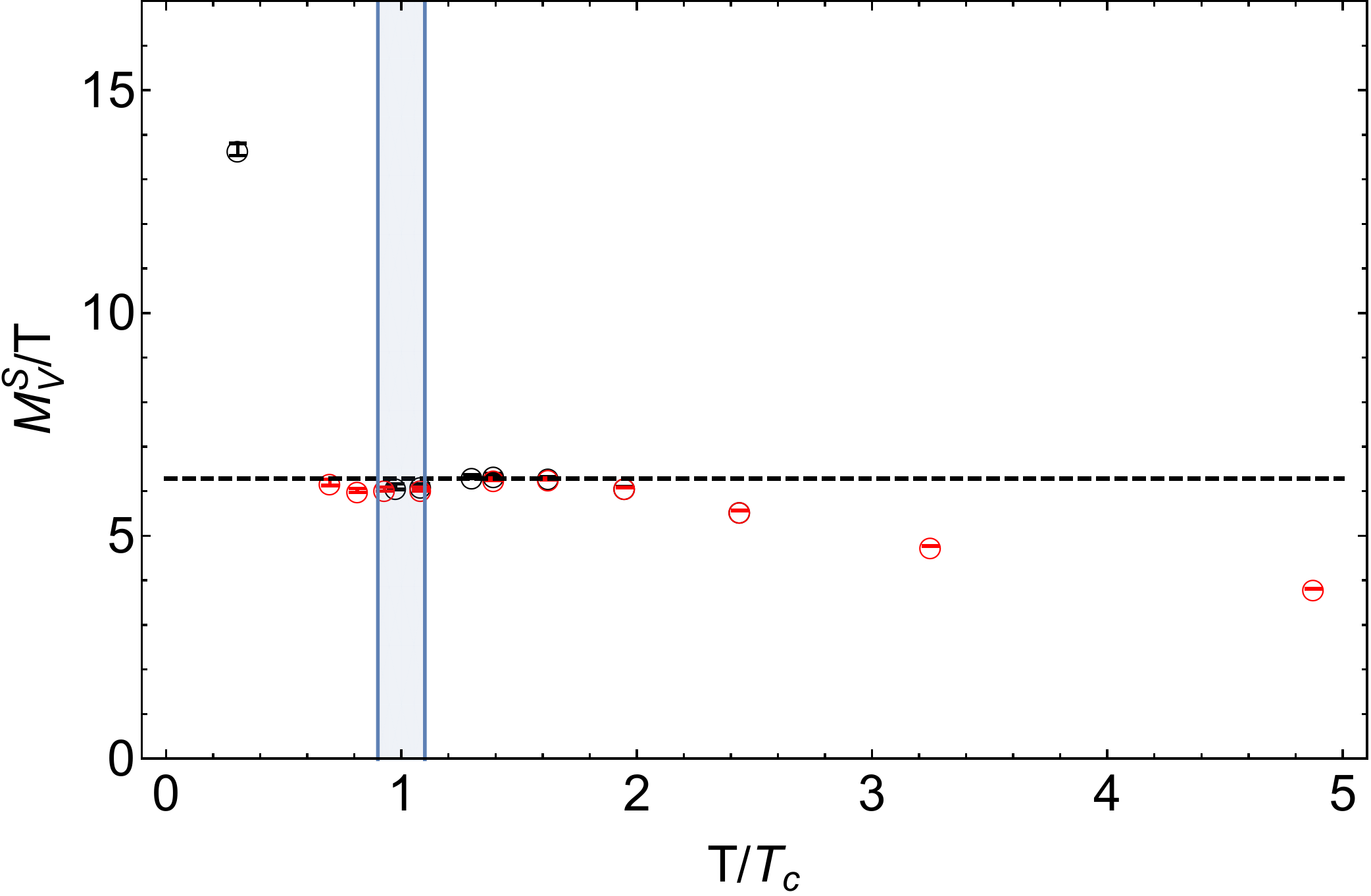}
\includegraphics[width=.85\textwidth]{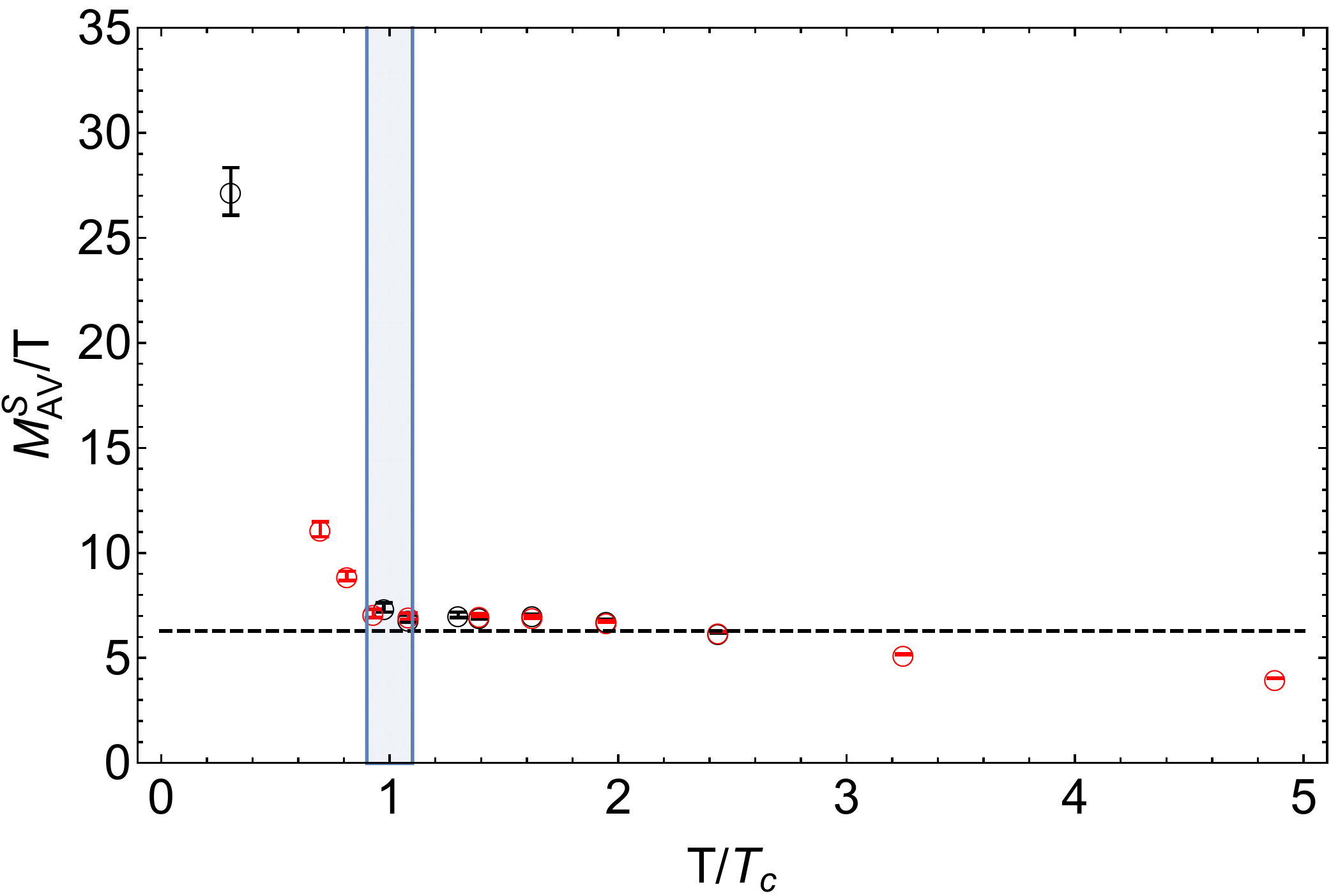}
\caption{%
\label{fig:smass_T_vector}%
Screening masses normalized by temperature for vector (top) and axial vector (bottom) mesons. 
Black and red points are extracted from ensembles of $N_t\times16^3$ and $N_t\times 16^2\times24$, respectively. 
The temperature is in units of  $T_c$, 
where the blue vertical band denotes the uncertainty of $T_c$ itself. 
}
\end{center}
\end{figure}

\begin{figure}
\begin{center}
\includegraphics[width=.85\textwidth]{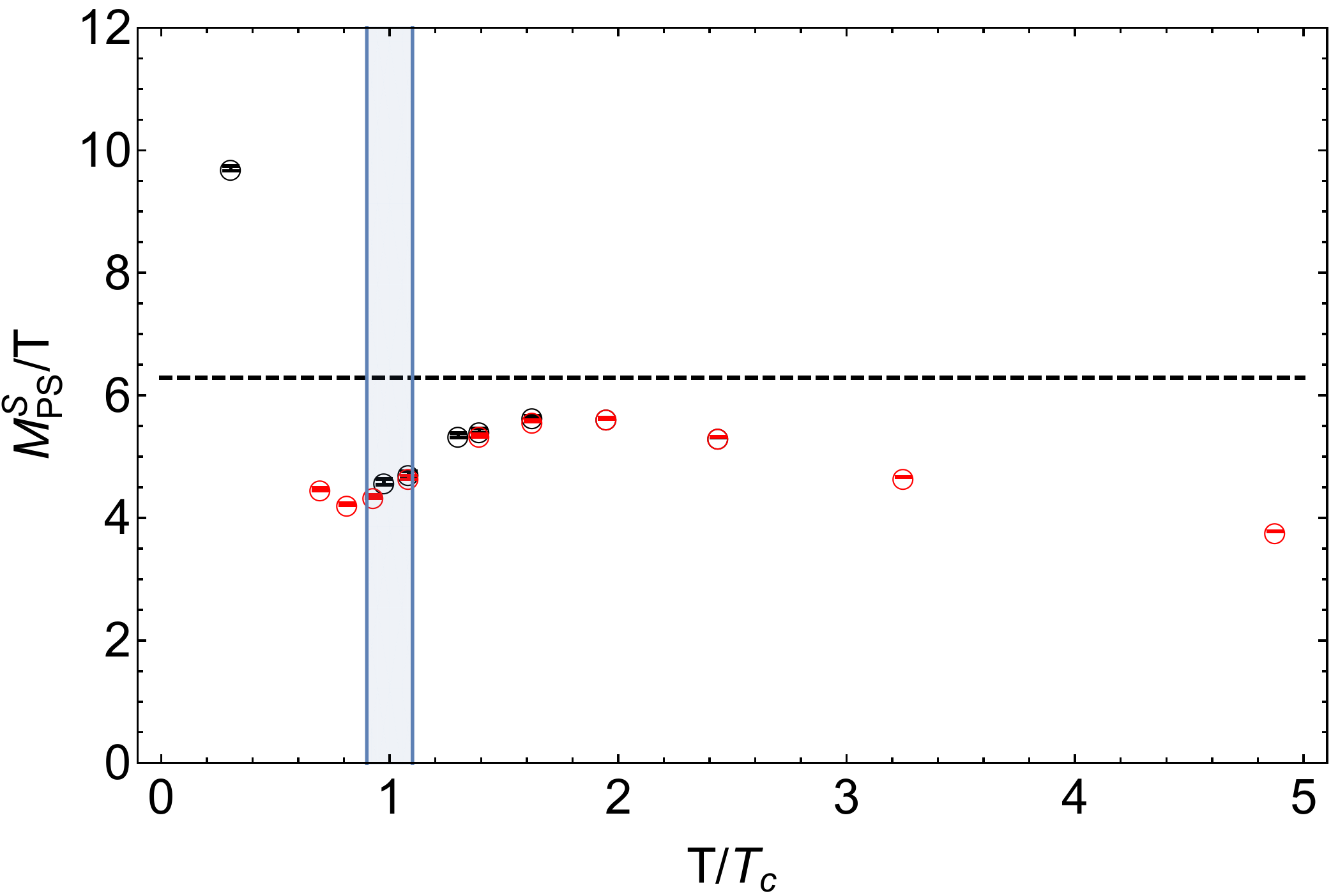}
\includegraphics[width=.85\textwidth]{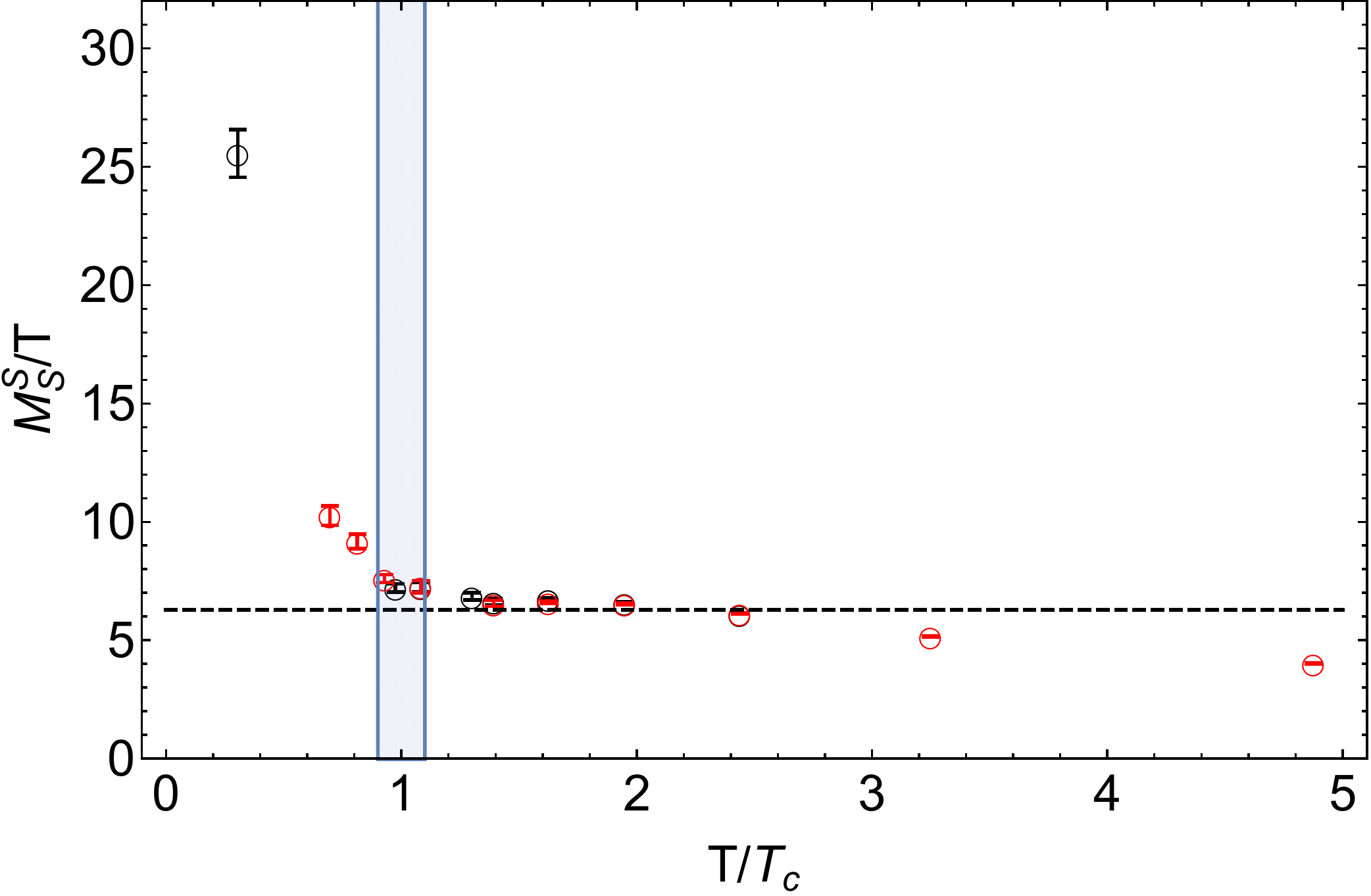}
\caption{%
\label{fig:smass_T_scalar}%
Screening masses normalized by temperature for pseudoscalar (top) and scalar (bottom) mesons. 
Black and red points are extracted from ensembles of $N_t\times16^3$ and $N_t\times 16^2\times24$, respectively. 
The temperature is in units of  $T_c$, 
where the blue vertical band denotes the uncertainty of $T_c$. 
}
\end{center}
\end{figure}

In the finite temperature calculations, it is often suggested to plot the screening mass divided by 
the temperature as it shows linear dependency above $T_c$. 
The results are shown in \Fig{smass_T_vector} and \Fig{smass_T_scalar}. 
The black dashed line corresponds to $2\pi$ which is associated with the Matsubara frequency 
for massless free quarks. 
For all mesonic channels, data points approach the dashed line as the temperature increases and 
seem to form a plateau. However, they start to deviate from the plateau above $2T_c$, possibly as a consequence of 
the finite lattice spacing.~\footnote{
As shown in $2+1$ Lattice QCD calculations using staggered fermions \cite{Cheng}, 
the size of these lattice artefacts can significantly be reduced if highly improved lattice fermions being used. 
} 
This suggests that in looking at $R_V$ and $R_S$ (and in general in discussing parity-doubling) one should not include in the physical 
very high temperatures, but rather restrict attention to $ T \lsim 2\,T_c$.

\section{Discussion}
\label{Sec:discussion}

We collected numerical evidence of the fact that the high-temperature behavior of the $SU(2)$ theory with $N_f=2$ Dirac 
fundamental fermions differs in three respects from
the low-temperature one. 
The numerical study of the Polyakov loop and its fit shows the existence of a pronounced peak in the  susceptibility. 
Its position identifies a temperature $T_c$, that we interpret in terms of the deconfinement 
(cross-over) temperature.
While the study of the details of the transition would require a dedicated program, this result 
is accurate enough to allow us to clearly separate the high-$T$ and low-$T$ regimes, and concentrate on the 
symmetry properties of the physical spectrum above $T_c$.

The study of temporal correlation functions shows that for $T>T_c$ the vector and axial-vector 2-point functions have
compatible  $t$-dependence, supporting the hypothesis that
 parity doubling is emerging at $T_c$, and global symmetry is restored.
 This is confirmed by the study of spatial correlation functions, in which one clearly sees that the behavior of the 
 screening masses of vectors and axial-vector mesons changes at $T\sim T_c$: while the two masses are different and depend 
 on $T$ in two different ways when $T<T_c$, for $T>T_c$ the masses come close to one another, and, most importantly,  
 show the same $T$-dependence.
 
 This last observation  suggests that the small splitting in the masses we observe is due to a combination of lattice artefacts
 (in particular finite spacing and finite quark mass).  To confirm or disprove this statement,
one would need to extend the study in this paper, and consider more than one value of the bare coupling and of the bare quark mass,
in order to extrapolate them both to the physically relevant regime.
By doing so, one might not only  be able to show that the mass difference between vectors and axial-vectors vanishes,
but also to study other properties of the transition itself, such as its order.

The numerical study of the scalar and pseudo-scalar masses, in which we focused on cleaner states that form a fundamental of $SO(5)$,
shows qualitative features that are in broad agreement with the restoration also of the axial $U(1)_A$ symmetry at high temperature.
Our data on spatial correlation functions 
seems to suggest that this is taking place at a larger temperature $T^{\prime}\sim 1.5 T_c$.
This is also supported by the fact that in the temporal correlation functions we do not see the effect of parity doubling in the spin-0 correlators,
for the same choice  $N_t=40$ for which the vector and axial-vector correlators do agree with one another.
This is the most striking element of novelty of this study, although it must be considered as preliminary.

This paper is to be understood as a first step in what is a potentially broad and extensive research program.
The results obtained are in good agreement with what expected on field-theory grounds about the 
non-trivial behavior of this theory at high temperatures: it deconfines, and both the global $SU(4)$ and axial $U(1)_A$ symmetries are 
restored.
Two main sets of explorations are interesting to pursue in the future.
On the one side, it is interesting to perform precision studies of this system, in which larger statistics,
and a broader set of values of the lattice parameters, are used in order to establish whether the three transitions we
identified are distinct (and in this case how to classify them, and precisely measure the critical temperatures),
or whether they are just three manifestations of the broader phenomenology related to a cross-over. 

On the other hand, it is also interesting to understand how the system reacts to
the introduction of additional sources of symmetry breaking at the Lagrangian level.
For example, the weak gauging of a subgroup of $Sp(4)$ (as in phenomenological composite-Higgs models),
is going to break the global symmetry of the model, and with it the large degeneracies of states.
It would be useful to know how these phenomena depend on finite temperature.
Closely related, although possibly simpler, is the question of what happens at finite $\mu$: given that $SU(2)$ is pseudo-real,
this model is free of the traditional sign problem of similar models with larger gauge groups.
It should hence be possible to attempt a more general study of the phase diagram as a function of both $T$ and $\mu$.

The richness of the field theory behavior of this model, the wide variety of its possible applications
 and the fact that this study shows that its thermal features are amenable to quantitative numerical studies, 
all contribute to making it an ideal environment in which to study  highly non-trivial phenomena, which might
 shed light on many  aspects of  direct relevance to QCD, TC and composite-Higgs scenarios.
In this paper we performed a first study along these lines, mainly aimed at collecting evidence of symmetry restoration
at high temperature. We also discussed ways to improve our results, and suggested 
avenues for further investigation, which we will pursue in the future.

\vspace{1.0cm}
\begin{acknowledgments}
\end{acknowledgments}

This work is supported in part by the STFC Consolidated Grant ST/L000369/1. 
The work of J.-W. L. is additionally supported by Korea Research Fellowship program 
funded by the Ministry of Science, ICT and Future Planning through 
the National Research Foundation of Korea (2016HID3A1909283). 
The authors thank S. Hands, G. Aarts, B. Jager, F. Attanasio and E. Bennett for discussions. 
Numerical computations were executed in part on the HPC Wales systems, 
supported by the ERDF through the WEFO (part of the Welsh Government). 
\vspace{1.0cm}
\appendix
\section{Spinors and global symmetries.}
\label{Sec:A}

We summarise in this Appendix  some useful notation about spinors, and show explicitly the origin of the enhanced 
global $SU(4)$ symmetry of the model.

The space-time metric is $\eta_{\mu\nu}\equiv{\rm diag}\,\{1,-1,-1,-1\}\,=\,\eta^{\mu\nu}$, and the Dirac algebra
is defined by the relation $\{\gamma^{\mu},\gamma^{\nu}\}=2\eta^{\mu\nu}$, with $\gamma^0$ hermitian and $\gamma^i$ anti-hermitian,
such that $\gamma^0\gamma^{\mu}\gamma^0=\gamma^{\mu\,\dagger}$.
Chirality is related to  $\gamma_5\equiv i \gamma^0\gamma^1\gamma^2\gamma^3$,
the left-handed(LH) chiral projector is $P_L=\frac{1}{2}\left( \mathbb{I}_4+\gamma_5\right)$,
and a 4-components LH chiral spinor $Q_L$ obeys $P_L Q_L=Q_L$.
The charge-conjugation matrix $C=i\gamma^2\gamma^0$ obeys  $C \gamma_{\mu} C^{-1} = - \gamma^{\mu\,T}$ and $C^2=-\mathbb{I}_4=-CC^{\dagger}$. 
 
The chiral representation for the gamma matrices, in terms of the Pauli matrices $\tau^i$, is 
\beqs
\gamma^0&=&\left(\begin{array}{cc}
0 & \mathbb{I}_2 \cr
\mathbb{I}_2 & 0
\end{array}\right)\,,~~
\gamma^i\,=\,\left(\begin{array}{cc}
0 & -\tau^i \cr
\tau^i & 0
\end{array}\right)\,,~~
\gamma^5\,=\,\left(\begin{array}{cc}
\mathbb{I}_2 & 0 \cr
0& -\mathbb{I}_2
\end{array}\right)\,,~~
C\,=\,\left(\begin{array}{cc}
 -i\tau^2 & 0 \cr
0 & i \tau^2 
\end{array}\right)\,.
\eeqs
The following is immediate:
\beqs
\gamma^0\gamma^{\mu}&=&
\left(\begin{array}{cc} \bar{\sigma}^{\mu} & 0 \cr 0 & \sigma^{\mu}\end{array}\right)\,,
~~~
C \gamma^0\gamma^{\mu} C^{-1}\,=\,
\left(\begin{array}{cc} {\sigma}^{\mu} & 0 \cr 0 & \bar{\sigma}^{\mu}\end{array}\right)\,,
\eeqs
where $\sigma^{\mu}=(1,-\tau^i)$ and $\bar{\sigma}^{\mu}=(1,\tau^i)$.

A Majorana spinor $\lambda$ obeys $\lambda=\pm\lambda_C\equiv\pm C\bar{\lambda}^T=
 \pm C\gamma^0 \lambda^{\ast}=\pm i\gamma^2\lambda^{\ast}$.
 We resolve the $\pm$ ambiguity by conventionally choosing the $+$ sign.
 
 Given a 2-component spinor $u$ we can build a 4-component Majorana spinor as
\beqs
\lambda&=&\left(\begin{array}{c} u\cr i\tau^2 u^{\ast}\equiv-\tilde{C}u^{\ast}\end{array}\right)\,,
\eeqs
so that  $\lambda=\lambda_C=\lambda_L+\lambda_R$, where
\beqs
\lambda_L&=&\left(\begin{array}{c} u\cr 0\end{array}\right)\,,~~~~~
\lambda_R\,=\,\left(\begin{array}{c} 0 \cr i\tau^2 u^{\ast}\equiv-\tilde{C}u^{\ast}\end{array}\right)\,.
\eeqs
In 4-component notation this ensures that $\lambda_L=C\overline{\lambda_R}^{\,\,T}$ and $\overline{\lambda_L}=\lambda_R^{\,\,T} C=-\lambda_R^{\,\,T} C^{-1}$.

With these definitions in place, and making use of the fact that Grassmann variables anticommute, after some algebra one finds that
\beq
i\,\overline{\lambda_R}\gamma^{\mu}\partial_{\mu}\lambda_R=
i\,\overline{\lambda_L}\gamma^{\mu}\partial_{\mu}\lambda_L\,+\,\frac{1}{2}\partial_{\mu}\left(
-i\,\overline{\lambda_L}\gamma^{\mu}\lambda_L
+i\,\overline{\lambda_R}\gamma^{\mu}\lambda_R
\right)\,,\label{Eq:identity}
\eeq
which implies that the kinetic term can be written equivalently in terms of $\lambda_R$ as of $\lambda_L$ (the total derivative can be dropped),
or equivalently one can write it in terms of the 4-component Majorana spinor $\lambda$ (with an overall factor of $\frac{1}{2}$ to avoid double counting).

We specify now the model of interest in this paper, with $SU(2)$ gauge symmetry.
Starting from the 2-component spinors $q^{i\,a}$, with $i=1\,,\,\cdots\,,\,4$ the flavor index and $a=1,2$ the color index,
we can build four LH and four right-handed(RH) 4-component spinors as
\beqs
q_L^{j\,a}&=&\left(\begin{array}{c} q^{j\,a} \cr 0 \end{array}\right)\,,~~~~~ q_R^{j\,a}\,=\,\epsilon^{ab}\left(\begin{array}{c} 0 \cr (-\tilde{C}q^{j\,\ast\,})_{b} \end{array}\right)\,,
\eeqs
with $j=1\,,\,\cdots\,,\,4$. Notice that the charge-conjugation used for the RH spinors implies to lower the $SU(2)$ indexes,
as it turns a fundamental of $SU(2)$ in its conjugate.
The essential property of $SU(2)$ is that this can be compensated by the $\epsilon^{ab}$ antisymmetric tensor. 
 
 One can define two Dirac spinors $Q^{i\,a}=q_{L}^{i\,a}+q_{R}^{i+2\,a}$, with $i=1,2$. 
 We identify such $Q^{i\,a}$ with the Dirac spinors that form the fundamental matter fields of the $SU(2)$ gauge theory.
Because of the structure of the gamma matrices, the kinetic terms do not couple different chiralities,
and hence we can write
\beqs
{\cal L}_K&=&\sum_{i=1,2}i\,\overline{Q^i}_{\,a}\gamma^{\mu}\,(D_{\mu}Q^i)^a\,=\sum_{i=1,2}\,\left(\frac{}{}i\,\overline{q^i_L}_{\,a}\gamma^{\mu}\,(D_{\mu}q^i_L)^a+i\,\overline{q^{i+2}_R}_{\,a}\gamma^{\mu}\,(D_{\mu}q^{i+2}_R)^a\right)\,,
\eeqs
which makes it immediately visible that there is a $U(N_f)^t\times U(N_f)^t=U(1)_A\times U(1)_B^t\times SU(2)_L^t\times SU(2)_R^t$ global symmetry,
as would be true in any $SU(N_c)$ gauge theory.

For $SU(2)$  the global symmetry is actually larger:
by making use of four LH 4-component spinors and of Eq.~(\ref{Eq:identity}) 
one has
\beqs
i\,\overline{q^i_R}_{\,a}\gamma^{\mu}\,(D_{\mu}q^i_R)^a&=&
i\,\overline{q^i_L}_{\,a}\gamma^{\mu}\,(D_{\mu}q^i_L)^a\,,
\eeqs
and hence we can write
\beqs
{\cal L}_K&=&\sum_{i=1}^4\,i\,\overline{q^i_L}_{\,a}\gamma^{\mu}\,(D_{\mu}q^i_L)^a\,=\,
\sum_{i=1}^4\,i\,(q^{i\,\dagger})_{\,a}\bar{\sigma}^{\mu}\,(D_{\mu}q^i)^a\,,
\eeqs
which is manifestly $SU(4)\times U(1)_A$ invariant.
 Notice that besides the $SU(2)^t\times SU(2)^t$, the $SU(4)\sim SO(6)$ group includes also the $U(1)^t_B$ associated with baryon number.

 For completeness, we can explicitly verify that 
 \beqs
 {\cal L}_m&=&-m\overline{Q^i}_{\,a} Q^{i\,a}\,=\,-m\epsilon_{ab} q^{i+2\,a} \tilde{C} q^{i\,b}\,+\,{\rm h. c.}
 \,=\,-m \frac{1}{2}\epsilon_{ab}q^{n\,a\,T}\tilde{C} q^{m\,b}\,\Omega_{nm}\,+\,{\rm h. c.}\,,
 \eeqs
 where the fact that $\Omega$ is antisymmetric comes from the antisymmetric $\epsilon_{ab}$.
 This is a Majorana mass, with $M=m \,\Omega$, which breaks explicitly the symmetry to $Sp(4)$.
The $U(1)_B^t$ is a subgroup of $Sp(4)\sim SO(5)$, hence the spectrum of composite states cannot be classified 
in terms of baryon number, as mesons and baryons are in common 
$Sp(4)$ multiplets.
In the case one gauges the baryon number, then the symmetry would be explicitly broken back to the familiar $U(2)^2$.

\section{$SU(4)$ and $Sp(4)$ algebra.}
\label{Sec:B}

The $N^2-1=15$ generators  of $SU(4)$ are hermitean traceless $4\times 4$ complex matrices $T^A$.
The $Sp(4)$ subgroup is defined as the matrices that leave invariant the symplectic $\Omega$, 
which we write as
\beqs
\Omega&=&
\left(\begin{array}{cccc}
0 & 0 & 1 & 0\cr
0 & 0 & 0 & 1\cr
-1 & 0 & 0 & 0\cr
0 & -1 & 0 & 0\cr
\end{array}\right)\,.
\label{Eq:symplectic}
\eeqs
$Sp(4)$  is generated by the subset of $10$ generators of $SU(4)$ that obey the relation
\beqs
\Omega T^A+T^{A\,T} \Omega&=&0\,,~~~~~ {\rm for}\, A\,=\,6\,,\,\cdots\,15\,,
\eeqs
while the $5$ broken generators  obey
\beqs
\Omega T^A-T^{A\,T} \Omega&=&0\,,~~~~~ {\rm for}\, A\,=\,1\,,\,\cdots\,5\,.
\eeqs

By imposing the normalization $\Tr T^AT^B=\frac{1}{2}\delta^{AB}$, we write the $15$ matrices as follows.
\beqs
\label{eq:su4_generators}
T^1&=&\frac{1}{2\sqrt{2}}
\left(
\begin{array}{cccc}
 0 & 1 & 0 & 0 \\
 1 & 0 & 0 & 0 \\
 0 & 0 & 0 & 1 \\
 0 & 0 & 1 & 0
\end{array}
\right)
\,,~~~~~
T^2\,=\,\frac{1}{2\sqrt{2}}
\left(
\begin{array}{cccc}
 0 & -i & 0 & 0 \\
 i & 0 & 0 & 0 \\
 0 & 0 & 0 & i \\
 0 & 0 & -i & 0
\end{array}
\right)
\,,~~~~~
T^3\,=\,\frac{1}{2\sqrt{2}}
\left(
\begin{array}{cccc}
 1 & 0 & 0 & 0 \\
 0 & -1 & 0 & 0 \\
 0 & 0 & 1 & 0 \\
 0 & 0 & 0 & -1
\end{array}
\right)
\,,\nonumber\\
T^4&=&\frac{1}{2\sqrt{2}}
\left(
\begin{array}{cccc}
 0 & 0 & 0 & -i \\
 0 & 0 & i & 0 \\
 0 & -i & 0 & 0 \\
 i & 0 & 0 & 0
\end{array}
\right)
\,,~~~~~
T^5\,=\,\frac{1}{2\sqrt{2}}
\left(
\begin{array}{cccc}
 0 & 0 & 0 & 1 \\
 0 & 0 & -1 & 0 \\
 0 & -1 & 0 & 0 \\
 1 & 0 & 0 & 0
\end{array}
\right)
\,,~~~~~
T^6\,=\,\frac{1}{2\sqrt{2}}
\left(
\begin{array}{cccc}
 0 & 0 & -i & 0 \\
 0 & 0 & 0 & -i \\
 i & 0 & 0 & 0 \\
 0 & i & 0 & 0
\end{array}
\right)
\,,\nonumber\\
T^7&=&\frac{1}{2\sqrt{2}}
\left(
\begin{array}{cccc}
 0 & 0 & 0 & -i \\
 0 & 0 & -i & 0 \\
 0 & i & 0 & 0 \\
 i & 0 & 0 & 0
\end{array}
\right)
\,,~~~~~
T^8\,=\,\frac{1}{2\sqrt{2}}
\left(
\begin{array}{cccc}
 0 & -i & 0 & 0 \\
 i & 0 & 0 & 0 \\
 0 & 0 & 0 & -i \\
 0 & 0 & i & 0
\end{array}
\right)
\,,~~~~~
T^9\,=\,\frac{1}{2\sqrt{2}}
\left(
\begin{array}{cccc}
 0 & 0 & -i & 0 \\
 0 & 0 & 0 & i \\
 i & 0 & 0 & 0 \\
 0 & -i & 0 & 0
\end{array}
\right)
\,,\nonumber\\
T^{10}&=&\frac{1}{2}
\left(
\begin{array}{cccc}
 0 & 0 & 1 & 0 \\
 0 & 0 & 0 & 0 \\
 1 & 0 & 0 & 0 \\
 0 & 0 & 0 & 0
\end{array}
\right)
\,,~~~~~
T^{11}\,=\,\frac{1}{2\sqrt{2}}
\left(
\begin{array}{cccc}
 0 & 0 & 0 & 1 \\
 0 & 0 & 1 & 0 \\
 0 & 1 & 0 & 0 \\
 1 & 0 & 0 & 0
\end{array}
\right)
\,,~~~~~
T^{12}\,=\,\frac{1}{2}
\left(
\begin{array}{cccc}
 0 & 0 & 0 & 0 \\
 0 & 0 & 0 & 1 \\
 0 & 0 & 0 & 0 \\
 0 & 1 & 0 & 0
\end{array}
\right)
\,,\\
T^{13}&=&\frac{1}{2\sqrt{2}}
\left(
\begin{array}{cccc}
 0 & 1 & 0 & 0 \\
 1 & 0 & 0 & 0 \\
 0 & 0 & 0 & -1 \\
 0 & 0 & -1 & 0
\end{array}
\right)
\,,~~~~~
T^{14}\,=\,\frac{1}{2\sqrt{2}}
\left(
\begin{array}{cccc}
 1 & 0 & 0 & 0 \\
 0 & -1 & 0 & 0 \\
 0 & 0 & -1 & 0 \\
 0 & 0 & 0 & 1
\end{array}
\right)
\,,~~~~~
T^{15}\,=\,\frac{1}{2\sqrt{2}}
\left(
\begin{array}{cccc}
 1 & 0 & 0 & 0 \\
 0 & 1 & 0 & 0 \\
 0 & 0 & -1 & 0 \\
 0 & 0 & 0 & -1
\end{array}
\right)
\,.\nonumber
\eeqs
The $5$  Goldstone bosons can be written as $\pi(x)=\sum_{A=1}^5 \pi^A(x)T^A$, or explicitly as
\beqs
\pi(x)&\equiv&\frac{1}{2\sqrt{2}}
\left(
\begin{array}{cccc}
 \pi^3(x) & \pi^1(x)-i \pi^2(x) & 0 & \pi^5(x)-i \pi^4(x) \\
 \pi^1(x)+i \pi^2(x) & -\pi^3(x) & i \pi^4(x)-\pi^5(x) & 0 \\
 0 & -i \pi^4(x)-\pi^5(x) & \pi^3(x) & \pi^1(x)+i \pi^2(x) \\
 i \pi^4(x)+\pi^5(x) & 0 & \pi^1(x)-i \pi^2(x) & -\pi^3(x)
\end{array}
\right)\,.
\eeqs

The maximal $SO(4)\sim SU(2)_L\times SU(2)_R$ subgroup of the unbroken $Sp(4)$ can be chosen to be generated by
\beqs
T^{1}_L&=&\frac{1}{2}\left(\begin{array}{cccc}
0 & 0 & 1 & 0\cr
0 & 0 & 0 & 0\cr
1 & 0 & 0 & 0\cr
0 & 0 & 0 & 0\cr
\end{array}\right)\,,\,\,
T^{2}_L\,=\,\frac{1}{2}\left(\begin{array}{cccc}
0 & 0 & -i & 0\cr
0 & 0 & 0 & 0\cr
i & 0 & 0 & 0\cr
0 & 0 & 0 & 0\cr
\end{array}\right)\,,\,\,
T^{3}_L\,=\,\frac{1}{2}\left(\begin{array}{cccc}
1 & 0 & 0 & 0\cr
0 & 0  & 0 & 0\cr
0 & 0 & -1 & 0\cr
0 & 0 & 0 & 0\cr
\end{array}\right)\,,\\
T^{1}_R&=&\frac{1}{2}\left(\begin{array}{cccc}
0 & 0 & 0 & 0\cr
0 & 0 & 0 & 1\cr
0 & 0 & 0 & 0\cr
0 & 1 & 0 & 0\cr
\end{array}\right)\,,\,\,
T^{2}_R\,=\,\frac{1}{2}\left(\begin{array}{cccc}
0 & 0 & 0 & 0\cr
0 & 0 & 0 & -i\cr
0 & 0 & 0 & 0\cr
0 & i & 0 & 0\cr
\end{array}\right)\,,\,\,
T^{3}_R\,=\,\frac{1}{2}\left(\begin{array}{cccc}
0 & 0 & 0 & 0\cr
0 &1  & 0 & 0\cr
0 & 0 & 0 & 0\cr
0 & 0 & 0 & -1\cr
\end{array}\right)\,.
\eeqs
The $T_L$ generators satisfy the $SU(2)_L$ algebra $\left[ T_L^i\,,\,T_L^j\right]=i\epsilon^{ijk}\,T_L^k$,
and similarly $\left[ T_R^i\,,\,T_R^j\right]=i\epsilon^{ijk}\,T_R^k$, while $\left[T_L^A,T_R^B\right]=0$.
These generators being all unbroken (in a vacuum aligned with $\Omega$), this is the natural choice 
of embedding of the $SO(4)$ symmetries of the Higgs field
in the context of composite Higgs.

The same model can be used also to describe traditional technicolor. 
In this case, the embedding of the Standard Model symmetries 
is based on the natural choice of generators of $SO(4)^t\sim SU(2)^t_L\times SU(2)^t_R$
as follows:
\beqs
t^{1}_L&=&\frac{1}{2}\left(\begin{array}{cccc}
0 & 1 & 0 & 0\cr
1 & 0 & 0 & 0\cr
0 & 0 & 0 & 0\cr
0 & 0 & 0 & 0\cr
\end{array}\right)\,,\,\,
t^{2}_L\,=\,\frac{1}{2}\left(\begin{array}{cccc}
0 & -i & 0 & 0\cr
i & 0 & 0 & 0\cr
0 & 0 & 0 & 0\cr
0 & 0 & 0 & 0\cr
\end{array}\right)\,,\,\,
t^{3}_L\,=\,\frac{1}{2}\left(\begin{array}{cccc}
1 & 0 & 0 & 0\cr
0 & -1  & 0 & 0\cr
0 & 0 & 0 & 0\cr
0 & 0 & 0 & 0\cr
\end{array}\right)\,,\\
t^{1}_R&=&-\frac{1}{2}\left(\begin{array}{cccc}
0 & 0 & 0 & 0\cr
0 & 0 & 0 & 0\cr
0 & 0 & 0 & 1\cr
0 & 0 & 1 & 0\cr
\end{array}\right)\,,\,\,
t^{2}_R\,=\,-\frac{1}{2}\left(\begin{array}{cccc}
0 & 0 & 0 & 0\cr
0 & 0 & 0 & 0\cr
0 & 0 & 0 & -i\cr
0 & 0 & i & 0\cr
\end{array}\right)\,,\,\,
t^{3}_R\,=\,-\frac{1}{2}\left(\begin{array}{cccc}
0 & 0 & 0 & 0\cr
0 & 0  & 0 & 0\cr
0 & 0 & 1 & 0\cr
0 & 0 & 0 & -1\cr
\end{array}\right)\,.
\eeqs
In this case, one finds that (with the vacuum aligned with $\Omega$) the breaking $SU(2)^t_L\times SU(2)^t_R\rightarrow SU(2)_V^t$ emerges,
and the unbroken generators are $t^A_V=(t^A_L+t^{A\,T}_R)$, or explicitly:
\beqs
t^{1}_V&=&\frac{1}{2}\left(\begin{array}{cccc}
0 & 1 & 0 & 0\cr
1 & 0 & 0 & 0\cr
0 & 0 & 0 & -1\cr
0 & 0 & -1 & 0\cr
\end{array}\right)\,=\,\sqrt{2}T^{13}\,,~~~~~
t^{2}_V\,=\,\frac{1}{2}\left(\begin{array}{cccc}
0 & -i & 0 & 0\cr
i & 0 & 0 & 0\cr
0 & 0 & 0 & -i\cr
0 & 0 & i & 0\cr
\end{array}\right)\,=\,\sqrt{2}T^{8}\,,\,\,\nonumber\\
t^{3}_V&=&\frac{1}{2}\left(\begin{array}{cccc}
1 & 0 & 0 & 0\cr
0 & -1  & 0 & 0\cr
0 & 0 & -1 & 0\cr
0 & 0 & 0 & 1\cr
\end{array}\right)\,=\,\sqrt{2}T^{14}\,.
\eeqs
The normalization is $\Tr t^A_Vt^B_V=\delta^{AB}$, as in this case we are writing the generators in the bifundamental representation.

The unbroken $U(1)_B^t$ associated with baryon number is generated by $T^{15}=\frac{1}{\sqrt{2}}(T_L^3+T_R^3)$,
while the anomalous axial $U(1)_A$ is generated by
\beqs
T_A&=&
\frac{1}{2\sqrt{2}}
\left(
\begin{array}{cccc}
 1 & 0 & 0 & 0 \\
 0 & 1 & 0 & 0 \\
 0 & 0 & 1 & 0 \\
 0 & 0 & 0 & 1
\end{array}
\right)\,.
\eeqs

\section{Fit results of renormalized parameters}
\label{Sec:C}
\begin{figure}
\begin{center}
\includegraphics[width=.7\textwidth]{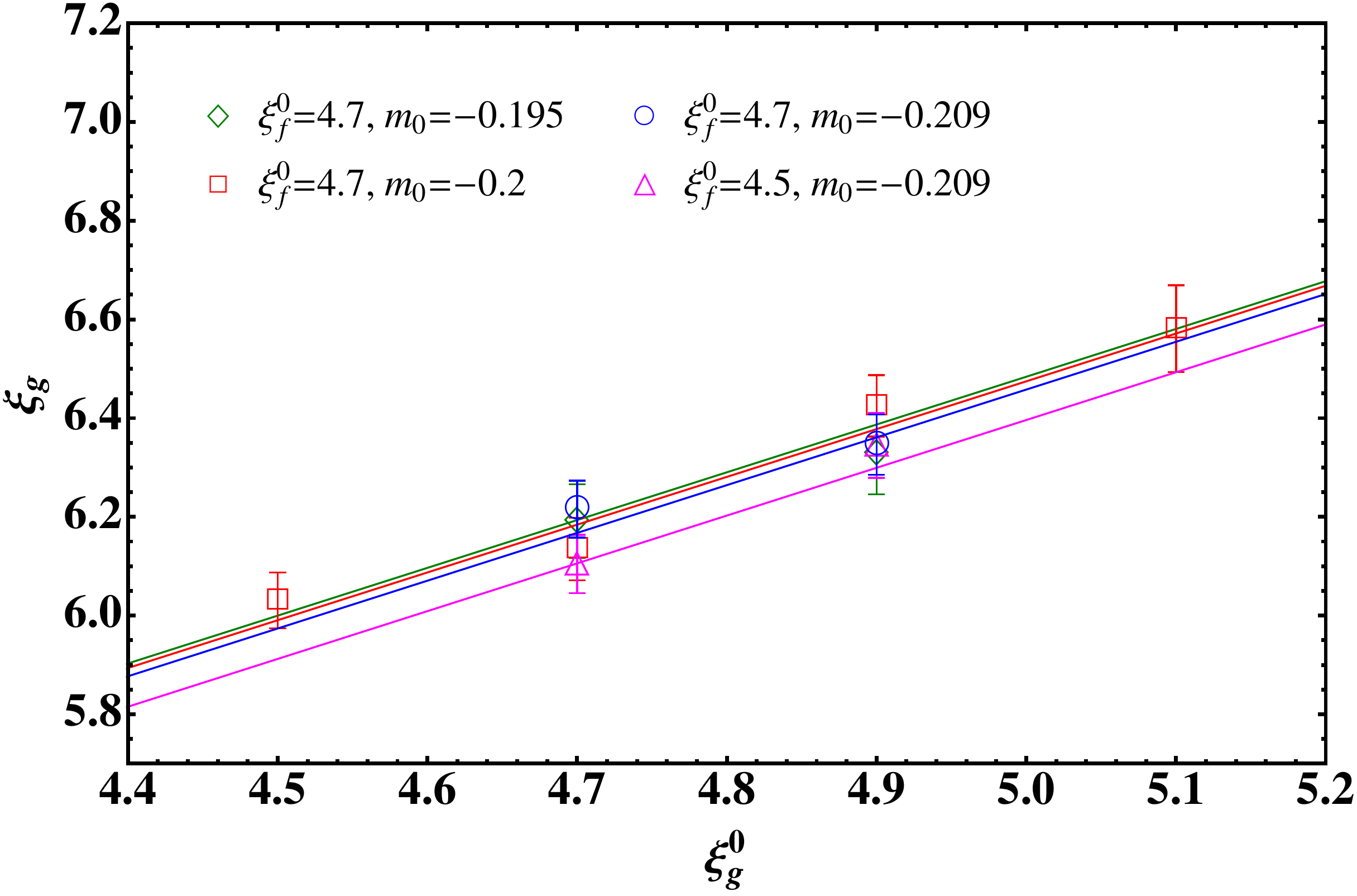}
\includegraphics[width=.7\textwidth]{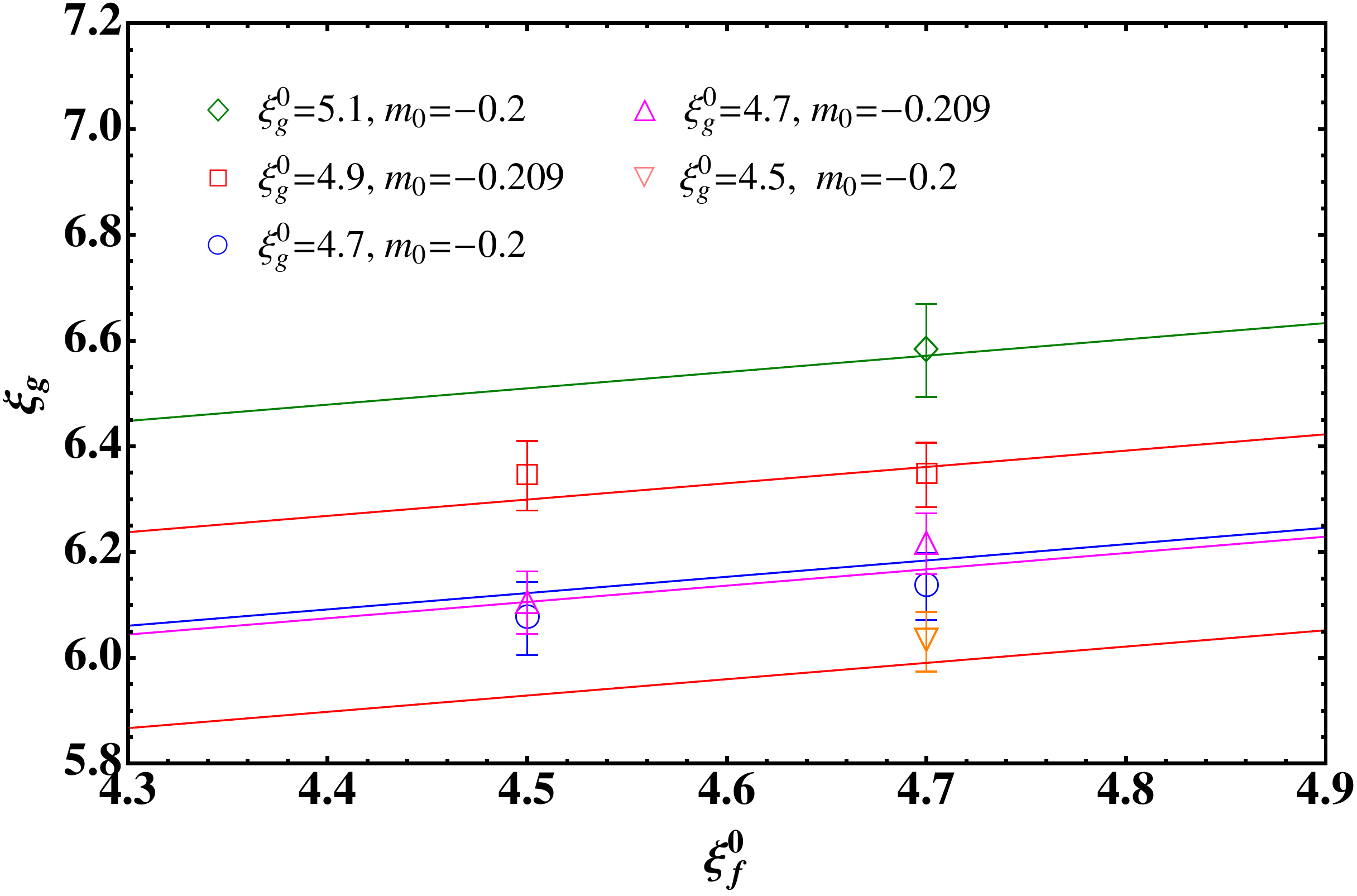}
\includegraphics[width=.7\textwidth]{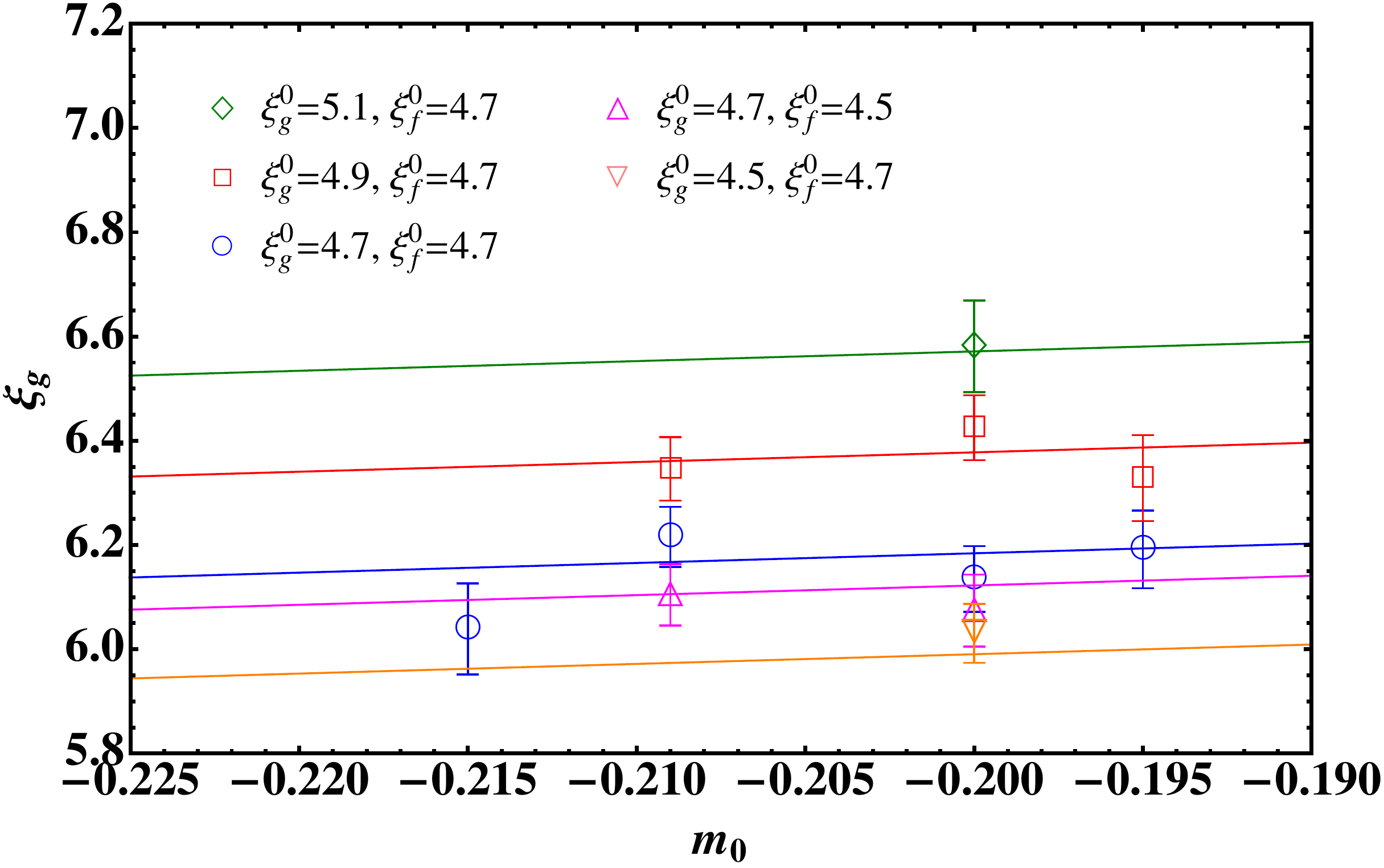}
\caption{%
\label{fig:xig_fit}%
Gauge anisotropy $\xi_g$ as a function of $\xi_g^0$, $\xi_f^0$, and $m_0$. 
The solid lines denote the fit functions in \Eq{renormalized_pm} 
where the coefficients are given by \Eq{fit_results}. 
}
\end{center}
\end{figure}

\begin{figure}
\begin{center}
\includegraphics[width=.7\textwidth]{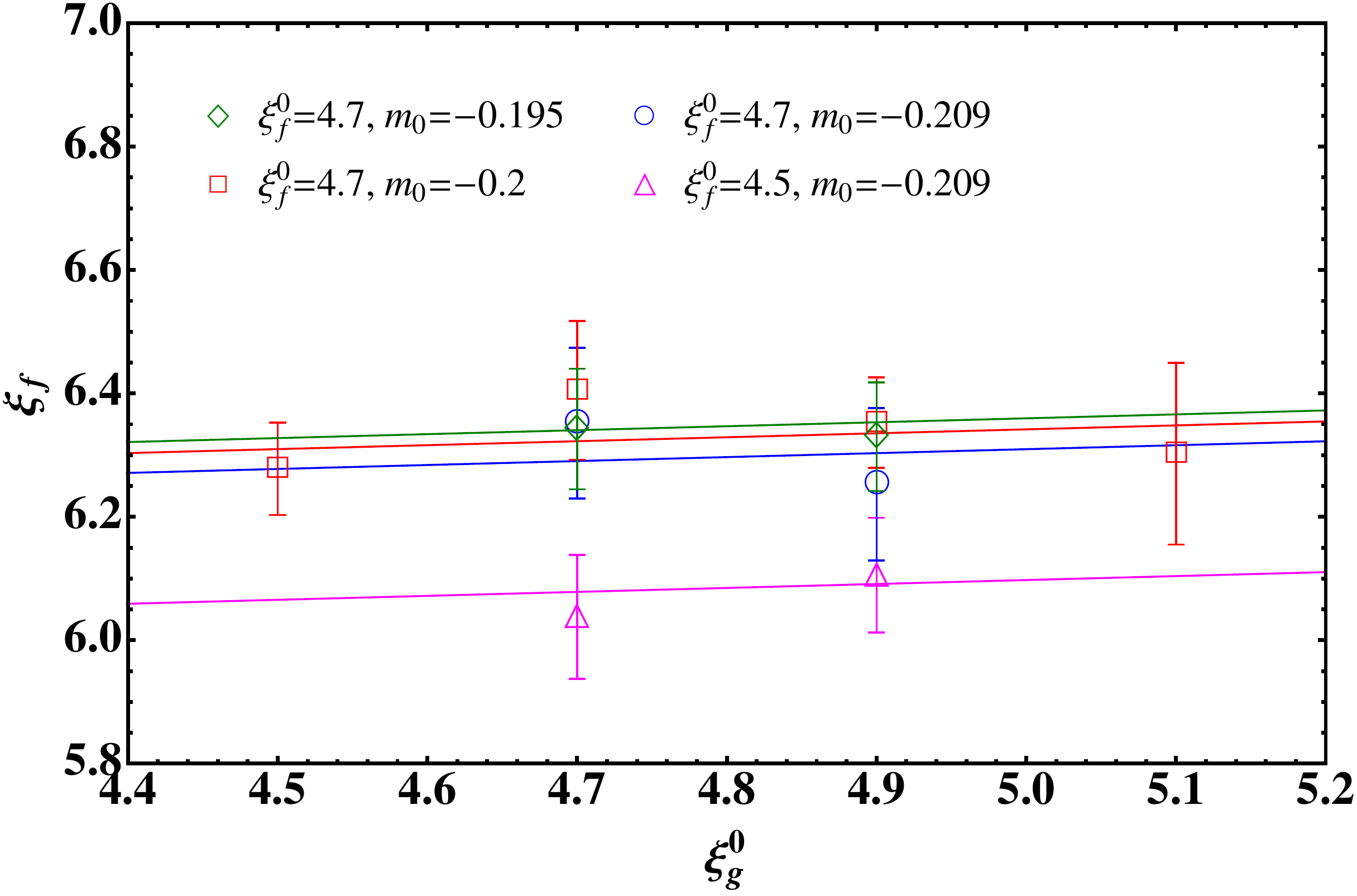}
\includegraphics[width=.7\textwidth]{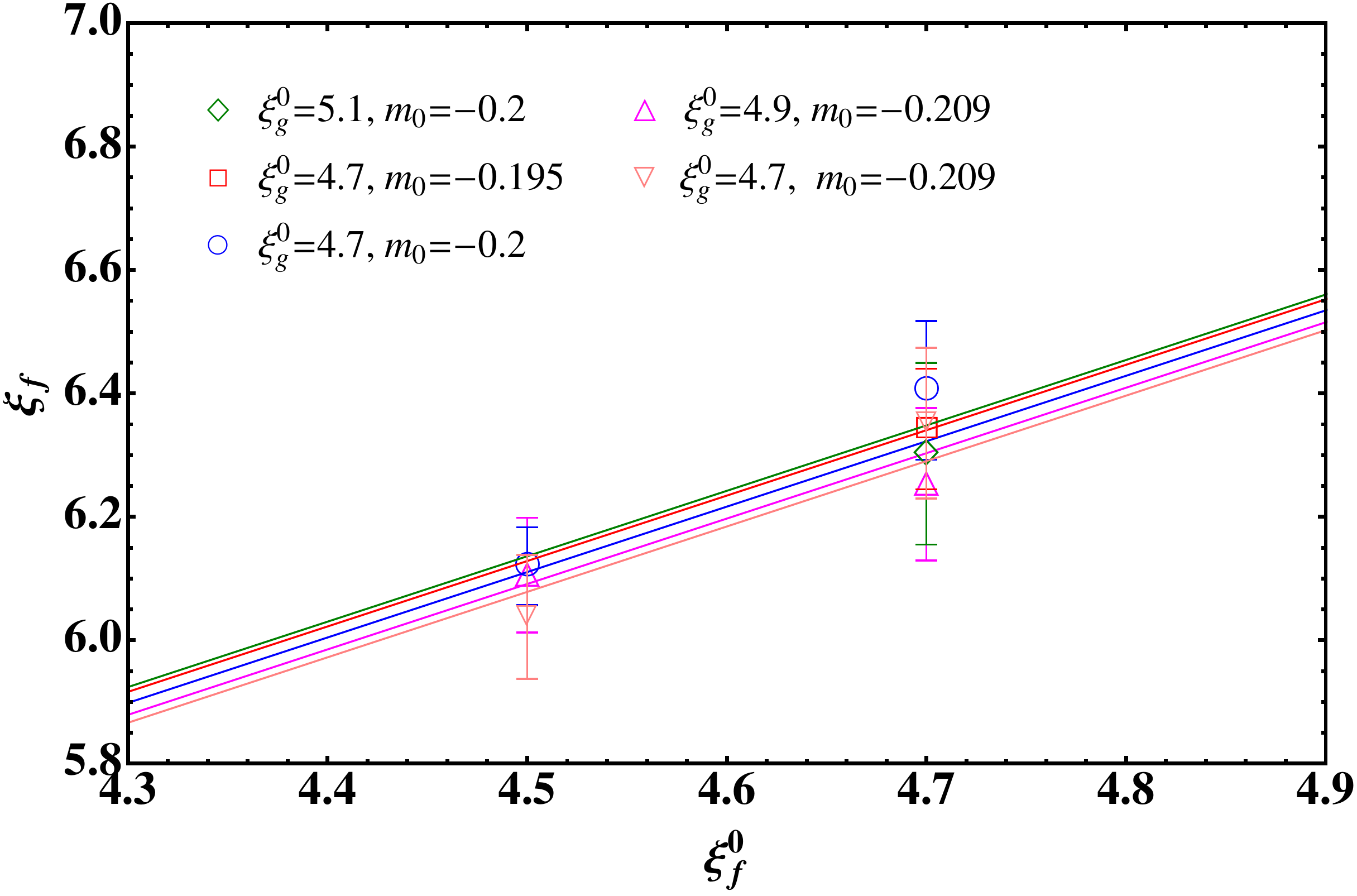}
\includegraphics[width=.7\textwidth]{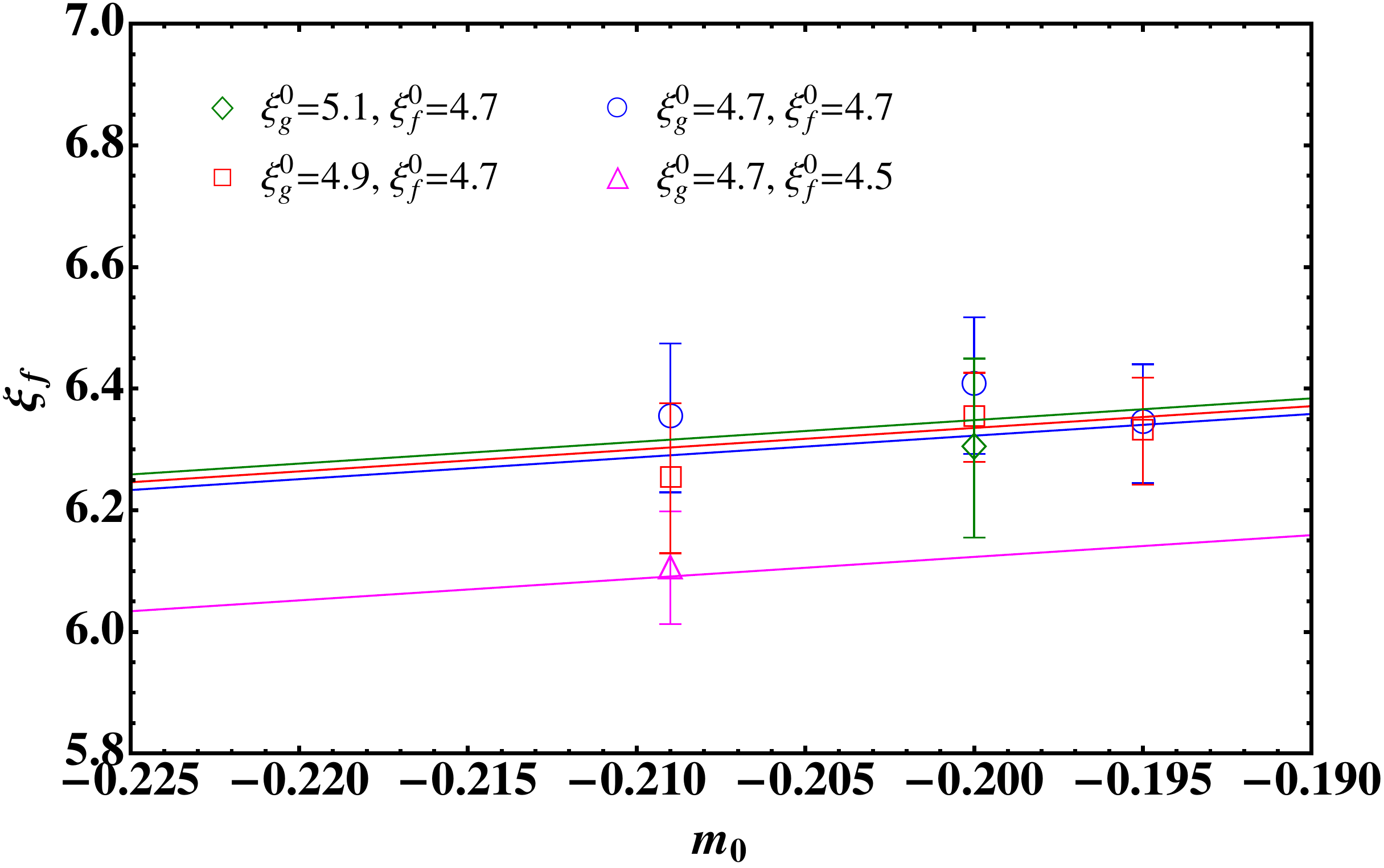}
\caption{%
\label{fig:xif_fit}%
Fermion anisotropy $\xi_f$ as a function of $\xi_g^0$, $\xi_f^0$, and $m_0$. 
The solid lines denote the fit functions in \Eq{renormalized_pm} 
where the coefficients are given by \Eq{fit_results}. 
}
\end{center}
\end{figure}

\begin{figure}
\begin{center}
\includegraphics[width=.7\textwidth]{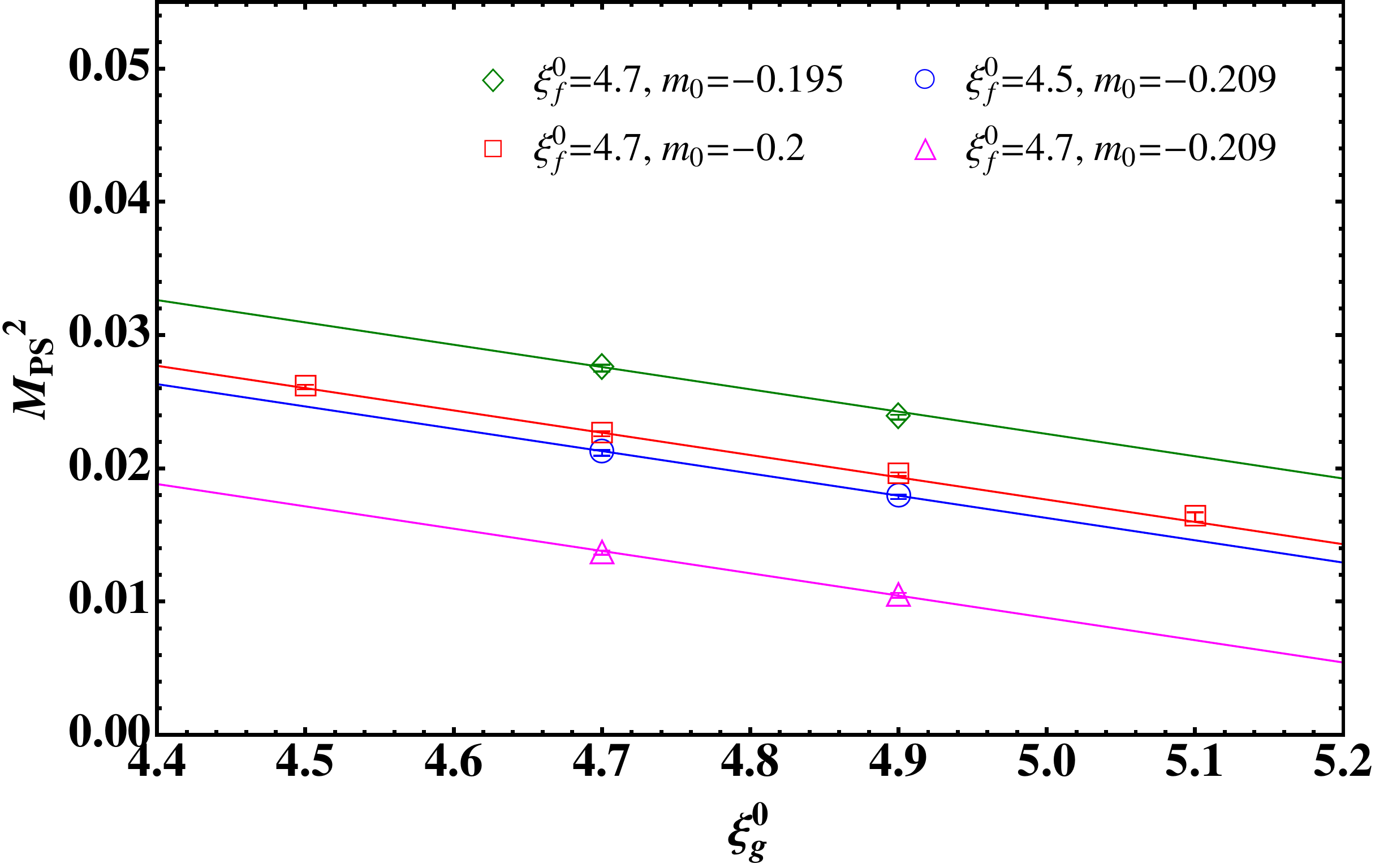}
\includegraphics[width=.7\textwidth]{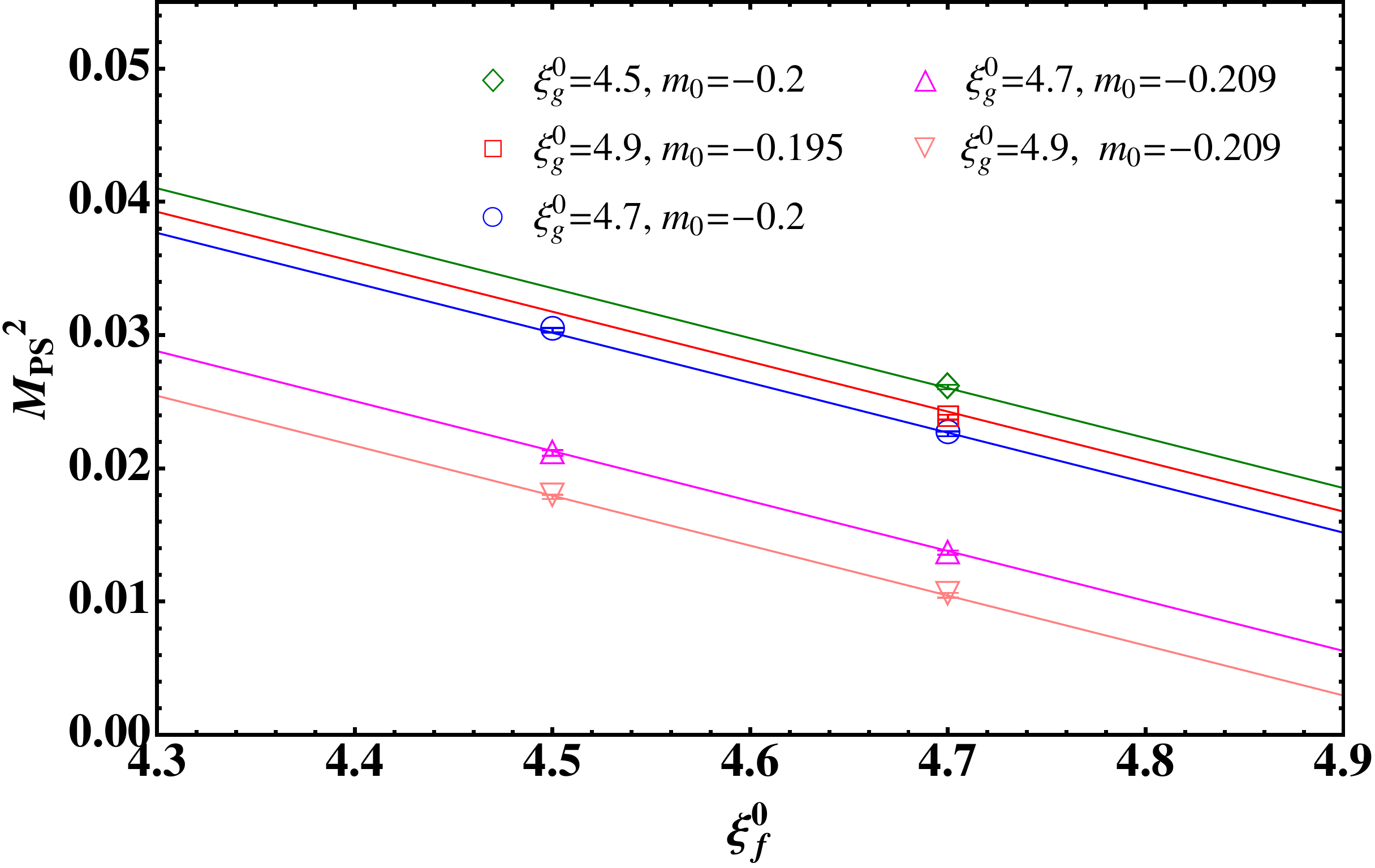}
\includegraphics[width=.7\textwidth]{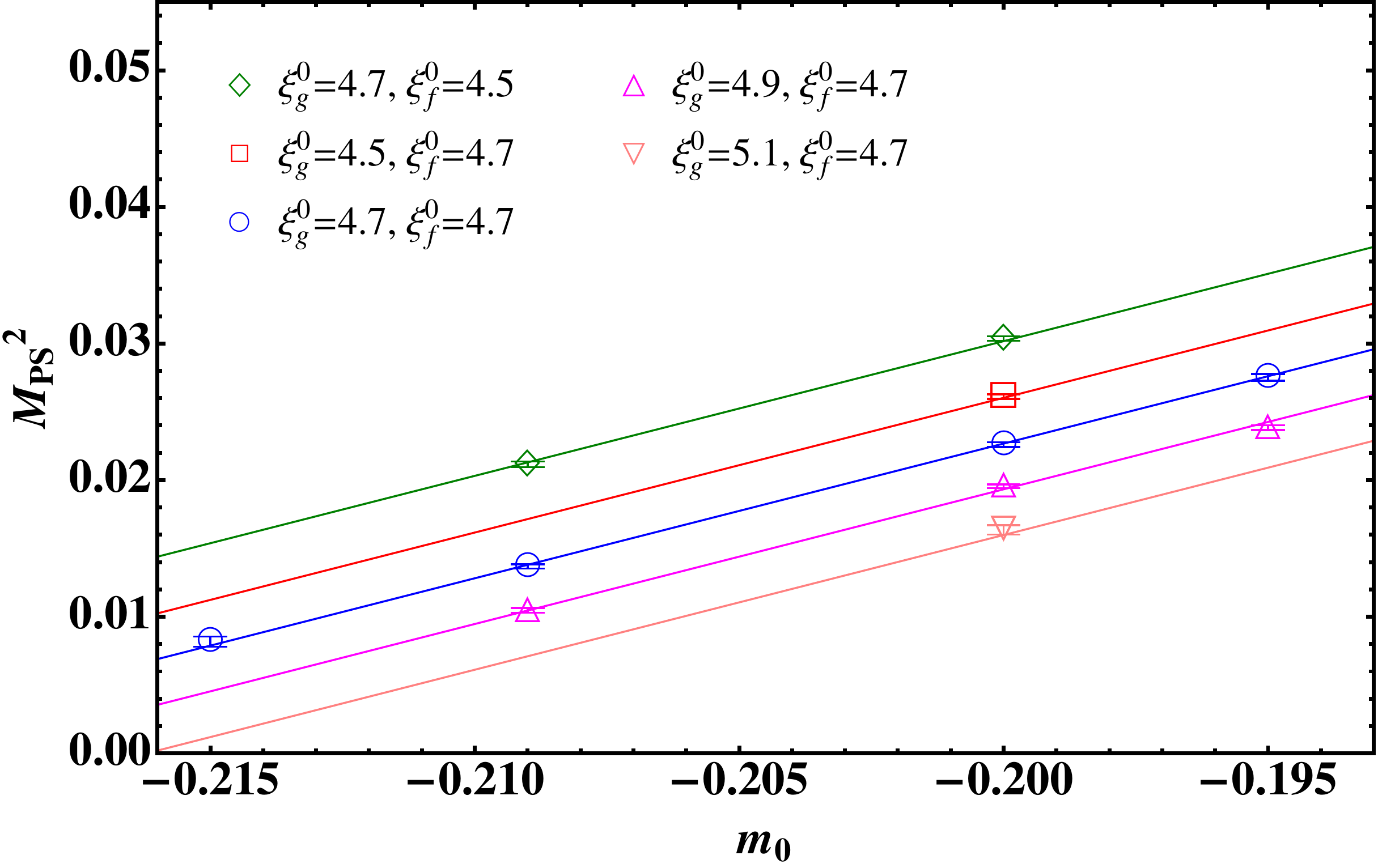}
\caption{%
\label{fig:m2_fit}%
Squared mass of a pseudoscalar meson $M_{PS}^2$ as a function of $\xi_g^0$, $\xi_f^0$, and $m_0$. 
The solid lines denote the fit functions in \Eq{renormalized_pm} 
where the coefficients are given by \Eq{fit_results}. 
}
\end{center}
\end{figure}

In this Appendix, we demonstrate how the fits in Section~\ref{Sec:tuning_results} work 
by showing the renormalized parameters in \Eq{renormalized_pm} 
along with the fit results of \Eq{fit_results} 
in the two-dimentional slices of the measured and lattice parameters. 
See \Fig{xig_fit} and \Fig{xif_fit}
for the fermion and gauge anisotropies, 
and see \Fig{m2_fit} for the squared mass of pseudoscalar meson $M_{PS}^2$. 

\end{document}